\documentclass[twocolumn]{aastex62}
\usepackage{subfigure}

\graphicspath{{./}{figures/}}

\submitjournal{ApJ}
\accepted{December 13. 2018}

\shorttitle{SZ observations of X-ray Cavities}
\shortauthors{Abdulla et. al.}
\begin{document}

\title{Constraints on the Thermal Contents of the X-ray Cavities of Cluster MS 0735.6+7421 \\ with Sunyaev-Zel'dovich Effect Observations}

\correspondingauthor{Zubair Abdulla}
\email{zabdulla@oddjob.uchicago.edu}

\author[0000-0002-0664-7812]{Zubair Abdulla}
\affil{Kavli Institute for Cosmological Physics, University of Chicago, 5640 South Ellis Avenue, Chicago, IL 60637, USA}
\affil{Department of Astronomy and Astrophysics, University of Chicago, 5640 South Ellis Avenue, Chicago, IL 60637, USA}

\author{John E. Carlstrom}
\affil{Kavli Institute for Cosmological Physics, University of Chicago, 5640 South Ellis Avenue, Chicago, IL 60637, USA}
\affil{Department of Astronomy and Astrophysics, University of Chicago, 5640 South Ellis Avenue, Chicago, IL 60637, USA}

\author{Adam B. Mantz}
\affil{Kavli Institute for Particle Astrophysics and Cosmology, Stanford University, 452 Lomita Mall, Stanford, CA 94305, USA}
\affil{Department of Physics, Stanford University, 382 Via Pueblo Mall, Stanford, CA 94305, USA}

\author{Daniel P. Marrone}
\affil{Steward Observatory, University of Arizona, 933 North Cherry Avenue, Tucson, AZ 85721, USA}

\author{Christopher H. Greer}
\affil{Steward Observatory, University of Arizona, 933 North Cherry Avenue, Tucson, AZ 85721, USA}

\author{James W. Lamb}
\affil{Owens Valley Radio Observatory, California Institute of Technology, Big Pine, CA 93513, USA}

\author{Erik M. Leitch}
\affil{Kavli Institute for Cosmological Physics, University of Chicago, 5640 South Ellis Avenue, Chicago, IL 60637, USA}
\affil{Department of Astronomy and Astrophysics, University of Chicago, 5640 South Ellis Avenue, Chicago, IL 60637, USA}

\author{Stephen Muchovej}
\affil{Owens Valley Radio Observatory, California Institute of Technology, Big Pine, CA 93513, USA}
\affil{Department of Astronomy, California Institute of Technology, Pasadena, CA 91125, USA}

\author{Christine O'Donnell}
\affil{Steward Observatory, University of Arizona, 933 North Cherry Avenue, Tucson, AZ 85721, USA}

\author{Thomas J. Plagge}
\affil{Kavli Institute for Particle Astrophysics and Cosmology, Stanford University, 452 Lomita Mall, Stanford, CA 94305, USA}
\affil{Department of Physics, Stanford University, 382 Via Pueblo Mall, Stanford, CA 94305, USA}

\author{David Woody}
\affil{Owens Valley Radio Observatory, California Institute of Technology, Big Pine, CA 93513, USA}

\begin{abstract}

Outbursts from active galactic nuclei (AGN) can inflate cavities in the intracluster medium (ICM) of galaxy clusters and are believed to play the primary role in offsetting radiative cooling in the ICM. However, the details of how the energy from AGN feedback thermalizes to heat the ICM is not well understood, partly due to the unknown composition and energetics of the cavities. The Sunyaev-Zel'dovich (SZ) effect, a measure of the integrated pressure along the line of sight, provides a means of measuring the thermal contents of the cavities, to discriminate between thermal, non-thermal, and other sources of pressure support. Here we report measurements of the SZ effect at 30 GHz towards the galaxy cluster MS 0735.6+7421 (MS0735), using the Combined Array for Research in Millimeter-wave Astronomy (CARMA). MS0735 hosts the most energetic AGN outburst known and lobes of radio synchrotron emission coincident with a pair of giant X-ray cavities $\sim 200$ kpc across. Our CARMA maps show a clear deficit in the SZ signal coincident with the X-ray identified cavities, when compared to a smooth X-ray derived pressure model. We find that the cavities have very little SZ-contributing material, suggesting that they are either supported by very diffuse thermal plasma with temperature in excess of hundreds of keV, or are not supported thermally. Our results represent the first detection (with $4.4 \sigma$ significance) of this phenomenon with the SZ effect.

\end{abstract}


\section{Introduction} \label{sec:intro}

Absent a source of heating, galaxy clusters with cool cores will quickly radiate away their available thermal energy via X-rays creating a reservoir of cool, dense gas ideal for star formation (reviewed in \citealt{Fabian1994CoolingGalaxies}). However, high resolution X-ray spectroscopy has not found evidence of the predicted cooling to low temperatures ($<1$ keV) \citep{Peterson2006X-rayClusters}, implying the radiative cooling is being counteracted by non-gravitational heating, most likely sourced by feedback from the central AGN (reviewed in \citealt{McNamara2007HeatingNuclei}). Outbursts from AGN can affect their surroundings through radiative feedback, where the radiation from the accreting super massive black hole couples directly to the cool gas at the center of clusters, and through radio-mechanical feedback, where jets driven by the AGN displace and heat the intracluster medium (ICM). Low frequency synchrotron radio emission produced by relativistic particles gyrating in magnetic fields generated by the AGN trace the path of the jets, which are often observed to terminate in extended radio lobes coincident with depressions in the X-ray surface brightness. The depressions are a result of the hot (typically $3-10$ keV) X-ray emitting ICM being displaced by the radio jets, creating ``cavities'' of lower density gas (reviewed in \citealt{Fabian2012ObservationalFeedback,Gitti2012EvidenceGroups}). The extents of X-ray cavities, which range from several to hundreds of kpc across, provide a gauge for the mechanical power output of radio-mechanical AGN feedback. X-ray observations of the cavities and radio observations of the outbursts have been combined to learn about the energetics of AGN outbursts and the role they play in heating the ICM; however, the dominant mechanisms by which the AGN feedback energy is converted to heat and then transferred to the ICM are still not well understood (reviewed in \citealt{McNamara2007HeatingNuclei,McNamara2012MechanicalClusters}).

The synchrotron emission from these lobes suggests they at least contain magnetic fields and a non-thermal distribution of relativistic electrons. However, estimates of the non-thermal pressure within the radio lobes, implied by equipartition of energies in the radiating relativistic electrons and magnetic fields, are smaller than that required to support the apparent bubbles against the pressure of the surrounding gas, as measured from X-ray data (e.g., \citealt{Fabian2001TheCluster,Blanton2003iChandra/i2052}). This indirectly suggests that heavy, non-radiating particles accompany the light, radiating particles and provide some pressure support, or alternatively that the jets are not near equipartition \citep{Dunn2004ParticleCores,DeYoung2006TheJets}. The heavy non-radiating particles may be supplied by the jets themselves or via entrainment of the ICM into the jets \citep{Croston2013TheClusters}. Constraints on the composition of the lobes would offer insight into how feedback heats the cluster atmosphere. Many channels for AGN heating have been put forward \citep{McNamara2007HeatingNuclei,McNamara2012MechanicalClusters}, including cavity heating \citep{Churazov2001EvolutionM87}, heating by shocks \citep{Fabian2003DeepRipples,Nulsen2005ClusterHydraA,Randall2015ADeepHistory} and sounds waves \citep{Fabian2006AConduction,Ruszkowski2004ClusterWaves} excited by the AGN jets, cosmic rays \citep{Lowenstein1991CosmicAnalysis,Guo2008FeedbackGalaxies,Ensslin2011CosmicHeating,Fujita2012NonClusters,Pfrommer2013TowardHESS,Weiner2013CosmicGalaxies,Ruszkowski2017CosmicMedium,Jacobs2017CosmicSolutions,Jacobs2017CosmicEmission}, and the mixing of thermal gas in bubbles with the surrounding ICM \citep{Hillel2017HitomiMixing,Hillel2017GentleClusters,Yang2016HowSimulations}. However, the relative balance of the various mechanisms remains poorly understood, in part because of the unknown composition of the cavities \citep{McNamara2012MechanicalClusters}.

If the cavities are in pressure balance with the surrounding ICM and supported by thermal plasma (e.g., entrained gas), the low density of the cavities implied by their X-ray surface brightness constrains this gas to be much hotter than the surrounding medium. Because the diffuse cavities are intrinsically faint, and are seen in projection with the brighter surrounding gas, X-ray spectroscopy alone can not rule out the possibility of very hot diffuse thermal gas in excess of tens of keV filling the cavities. In some systems, the missing X-ray surface brightness in the cavities has been used to constrain the temperature of thermal plasma potentially supporting the cavities to $kT > 20-50$ keV \citep{Nulsen2002InteractionA,Blanton2003iChandra/i2052,Sanders2007ARays}. Lacking a direct means of measuring the thermal contents of the cavities, the details of the composition of the cavities and relative balance of magnetic fields, non-thermal relativistic particles, and thermal plasma remain poorly understood. 

The SZ effect (\citealt{Sunyaev1972TheGalaxies}, see \citealt{Birkinshaw1998TheEffect,Carlstrom2002CosmologyEffect} for reviews), the inverse-Compton scattering of cosmic microwave background (CMB) photons off the energetic electrons of the ICM, is a measure of the integrated pressure along the line of sight, and so provides a unique tool for distinguishing between cavities supported by non-thermal or thermal plasma. A thermally supported cavity, even at very high temperatures, can provide a signal distinguishable from a cavity supported by non-thermal relativistic particles and magnetic fields, which contribute comparably minimally to the SZ Effect \citep{Pfrommer2005UnveilingEffect,Colafrancesco2005TheCavities,Prokhorov2010ComptonizationCocoons}. Only recently have advances in instrumentation made such measurements feasible.

We targeted MS 0735.6+7421 (hereafter MS0735) for observations of the SZ effect; this cluster has been extensively studied using X-ray data from XMM-\textit{Newton} \citep{Gitti2007Cosmological0735+7421} and \textit{Chandra} \citep{Vantyghem2014CyclingClusters} and radio data from the VLA at 327 MHz, 1.4 GHz, and 8.5 GHz \citep{Birzan2008RadiativePower}. These previous observations reveal a cool-core cluster with large X-ray cavities ($\sim 200$ kpc in diameter) coincident with radio lobes of synchrotron emission provided by the AGN-driven jet. From the extent of the cavities and pressure of the surrounding gas derived from the X-ray data along with an estimate of the cavity age \citep{Birzan2004AGalaxies}, the power to inflate the bubbles is estimated to be $1.7 \times 10^{46}$ erg s$^{-1}$, making it the most powerful AGN outburst known \citep{Vantyghem2014CyclingClusters}. The cavities are surrounded by X-ray bright rims of cool gas, which are measured to be in pressure balance with the surrounding ambient gas \citep{Vantyghem2014CyclingClusters}. The cavities are coincident with 327 MHz and 1.4 GHz radio lobe emission, but are lacking detected 8.5 GHz radio emission, classifying them as radio ghosts (\citealt{Birzan2008RadiativePower}, see section \ref{sec:contamination}). Under the assumptions of equipartition and pressure equilibrium, the X-ray and radio data for MS0735 imply that the ratio of the energy in heavy, non-radiating particles to that in radiating electrons in the cavities is $\sim 1000$ \citep{Birzan2008RadiativePower}.

In this work, we present high-resolution 30 GHz observations obtained with CARMA of the SZ effect from MS0735 to probe the composition of its giant X-ray cavities. We describe the observations, reductions, and map-making procedures for the CARMA data in section \ref{sec:obs}; discuss the task of building an appropriate model for the observed cluster components in sections \ref{sec:anal}; and present the results and conclusions of our analysis in sections \ref{sec:results} and \ref{sec:conclusion}, respectively. MS0735 has a redshift of $z=0.216$, corresponding to a scale of 3.53 kpc/\arcsec, (assuming a $\Lambda$CDM cosmological model with $\Omega_\Lambda = 0.7$, $\Omega_m = 0.3$, and $H_0 = 70 $km s$^{-1}$ Mpc$^{-1}$).

\section{Observations} \label{sec:obs}

CARMA\footnote{Decommissioned in April, 2015; https://www.mmarray.org/} was a 23 element interferometer that consisted of six 10.4 m, nine 6.1 m, and eight 3.5 m telescopes, which provided fields of view of FWHM 3.8\arcmin, 6.6\arcmin, and 11.4\arcmin, respectively, at 30 GHz. All 23 telescopes were capable of 30 GHz and 90 GHz observations. CARMA had two correlators, an 8-station ``wide-band'' (WB) correlator with 8 GHz bandwidth per baseline, and a more flexible ``spectral-line'' (SL) correlator, which could be configured to 15 stations of up to 8 GHz bandwidth or, as for these observations, 23 stations of 2 GHz bandwidth per baseline. In the 23-element observations, both correlators were used simultaneously to provide data for all possible baselines (using the SL correlator), while maximizing the sensitivity of selected baselines (using the WB correlator). Examples of previously presented 23-element CARMA observations of the SZ effect can be found in \cite{Brodwin2015THE1} and \cite{Mantz2017The122}.

In February--March of 2013, as part of the commissioning of the 23-element 30 GHz instrument, MS0735 was observed in a non-standard, compact configuration. In 2014, MS0735 was observed in CARMA-23 mode again, with the 10.4 m and 6.1 m telescopes arranged in the standard compact ``E'' configuration, and the 3.5 m telescopes in the ``SH'' configuration\footnote{In this configuration, six of the 3.5 m telescopes are arranged compactly while the other two ``outrigger'' stations provide longer baselines suitable for detecting and constraining point-like sources}. In both observation cycles, the eight most compactly arranged 6.1 m telescopes were directed to the WB correlator to maximize the sensitivity of the baselines corresponding to arcminute scales. The 23-element observations were two-pointing mosaics, with each pointing centered near an X-ray identified cavity to maximize sensitivity to those regions. In each 4--8 hour observation track, the array observed the source in 14 minute cycles, split equally between the two pointings, interspersed with observations of a calibrator. In addition, archival CARMA observations of MS0735 with the 8-element 3.5 m array in the ``SH'' configuration, pointed at the center of the cluster are included in our analysis. Details of these CARMA observations are summarized in Table \ref{table:observations}.\footnote{The raw CARMA data used in this work is available on the CARMA data archive: http://carma-server.ncsa.uiuc.edu:8181/. Our reduced data can be made available upon request.}

\begin{deluxetable*}{c c c c c c c c}[t]
\tablecaption{CARMA observations summary \label{table:observations}} 
\tablehead{Project ID & Observations Period & Array & Pointing Center & Integration time (hrs)\tablenotemark{a} & a $\arcsec$ & b $\arcsec$ & $\phi$ (deg)}
\startdata
c0876 & MAY-SEP 2012 & CARMA-8 & Cluster Center\tablenotemark{b} & 34.9 & 32.0 & 28.8 & 318 \\
\tableline
cx344 & FEB-MAR 2013 & CARMA-23 (WB) & SW cavity\tablenotemark{c} & 16.1 & 66.7 & 43.2 & 51 \\
- & - & - & NE cavity\tablenotemark{d} & 16.7 & 66.5 & 43.2 & 50 \\
- & - & CARMA-23 (SL) & SW cavity & 11.7 & 13.7 & 13.2 & 273 \\
- & - & - & NE cavity & 12.2 & 13.8 & 13.2 & 275 \\
\tableline
c1275 & JUN-OCT 2014 & CARMA-23 (WB) & SW cavity & 11.1 & 65.1 & 48.2 & 37 \\
- & - & - & NE cavity & 11.5 & 65.0 & 48.3 & 37 \\ 
- & - & CARMA-23 (SL) & SW cavity & 5.6 & 27.4 & 24.5 & 74 \\ 
- & - & - & NE cavity & 5.9 & 27.4 & 24.5 & 75 \\ 
\enddata
\tablenotetext{a}{Integration time = (\# of unflagged visibilities $\times$ individual integration time) / (\# of baselines $\times$ \# of 500 MHz frequency bands)}
\tablenotetext{b}{Cluster center: $07^\mathrm{h} 41^\mathrm{m} 44^\mathrm{s}, +74^{\circ} 14\arcmin 38\arcsec$}
\tablenotetext{c}{Southwest cavity: $07^\mathrm{h} 41^\mathrm{m} 49^\mathrm{s}, +74^{\circ} 15\arcmin 22\arcsec$}
\tablenotetext{d}{Northeast cavity: $07^\mathrm{h} 41^\mathrm{m} 39^\mathrm{s}, +74^{\circ} 13 \arcmin 51\arcsec$}
\tablecomments{A summary of the CARMA observations used in this analysis, split by era of observations, pointing center, and sub-array (see section \ref{sec:obs}). CARMA-8 refers to the 3.5 m 8-element observations, CARMA-23(WB) refers to the eight 6.1 m telescopes directed to the ``wide-band'' correlator in CARMA-23 observations, and CARMA-23(SL) refers to the remaining baselines directed to the ``spectral-line'' correlator in CARMA-23 observations. Observations were obtained in 4--8 hrs blocks with approximately 20--30\% overhead for calibration. Integration time represents the effective on-source integration time, which approximately accounts for flagged data. The synthesized (and composite in the case of CARMA-23(SL)) beam of each sub-array is described by the estimated semi-major axis $a$, semi-minor axis $b$ and rotation angle $\phi$. }
\end{deluxetable*}
The data reduction is done with a MATLAB based pipeline using the procedure described in \cite{Muchovej2007ObservationsArray}, which flags for weather, shadowing, poorly calibrated data, and technical issues; performs bandpass calibration using a bright quasar observed at the beginning of each observation track; and performs phase and gain calibration using a bright quasar observed at 15 minute intervals throughout each track. The 23-element SL and 8-element WB data are reduced as separate observations and overlapping baselines in the SL and WB data are flagged from the lower-bandwidth SL data. The absolute calibration is tied to periodic observations of Mars using the compact sub-array for which Mars is unresolved, bootstrapped to the entire array using a bright, unresolved quasar observed contemporaneously with the Mars observations. The absolute flux scale is set by comparing these observations to an update of the Mars model of \cite{Rudy1987Mars:Properties} that is accurate to 5 percent \citep{Perley2013ANGHz}. 

\subsection{Modeling} \label{sec:modeling}

Models for radio sources, ambient cluster gas, and cavities are fit simultaneously to all data in the $uv$-plane using a Markov chain Monte Carlo (MCMC) routine. The data are separated into unique baseline types and 500 MHz frequency bands (sixteen bands for CARMA-8 and CARMA-23 WB observations and four bands for CARMA-23 SL observations) when modeling. To create the Markov chain, candidate parameter values for the model are chosen from a broad parameter space, the likelihood of the model is computed, and the candidate values are either accepted or rejected based on the Metropolis-Hastings algorithm \citep{Metropolis1953EquationsMachines,Hastings1970,gilks1995markov}. The likelihood of a model is given by,
% With sufficiently large numbers, the Markov chain of accepted parameter values approaches the true probability distribution function (Gamermann 1997, Gilks, Richardson and Spiegelhalter 1996, MacKay 2003).
%
\begin{equation}
  \textit{L} = \prod_k exp \bigg(- \frac{1}{2}(\Delta R_k^2 + \Delta I_k^2) W_k \bigg),
  \label{eq:likelihood}
\end{equation}
where $\Delta R_k$ and $\Delta I_k$ are the differences between the model and the data for the real and imaginary components, respectively, at each visibility $k$ and the data weights $W_k=1/\sigma_k^2$ are the inverse variance of the interferometric visibilities (see \citealt{Reese2000Sunyaev0016+16,Bonamente2004MarkovData}). The same approach has been used in previous CARMA observations of galaxy clusters (e.g., \citealt{Plagge2013CARMAJ1347.51145,Mantz2014TheCARMA,Brodwin2015THE1}).

\subsection{Imaging} \label{sec:imaging}

These observations are assembled from a small mosaic of pointings made with a heterogeneous array, which complicates the process of making an interferometric image. Imaging of this interferometric data are done specifically for visualization, while all model fitting occurs in the $uv$-plane, where the noise of the data and the spatial response of the interferometer are well understood.

\begin{figure}[h]
  \centering
  \includegraphics[width=1.0\linewidth]{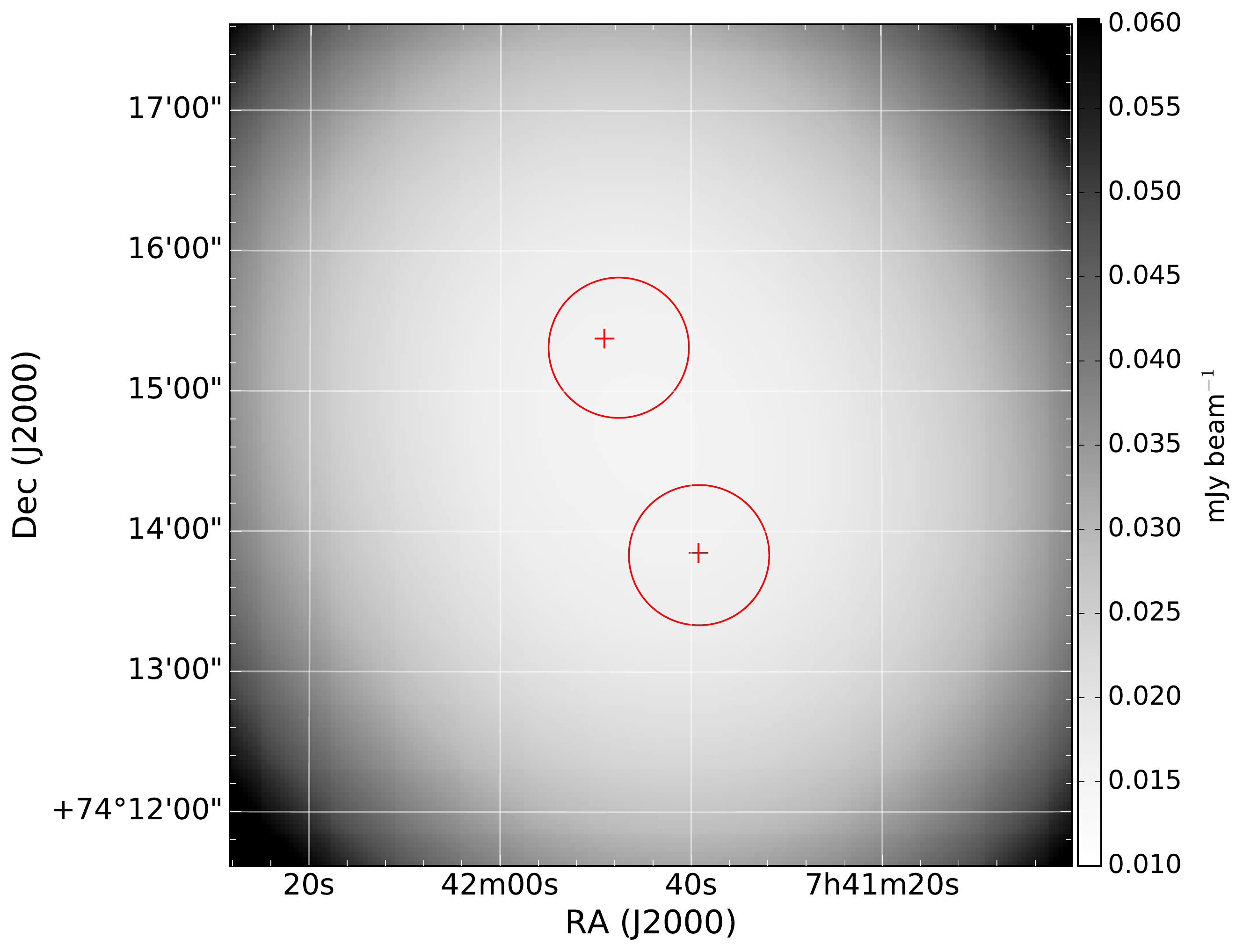}
  \caption{A noise map of our CARMA observations in mJy beam$^{-1}$, used to produce the CARMA SNR maps in this work. The positions of the two pointings of the CARMA-23 mosaic are shown with red crosses. The CARMA-8 pointing is at the map center. The X-ray identified cavities from \cite{Vantyghem2014CyclingClusters} are shown as red circles. Note the near uniform noise in the map throughout the regions of interest. At the center of the map, the noise $\sigma_\mathrm{rms}=0.0144$ mJy beam$^{-1}$.}
  \label{fig:ms0735_noise}
\end{figure}

We present maps in units of signal-to-noise ratio (SNR) to account for the non-uniform noise level in our maps owing to the unique primary beams of the heterogeneous array. For map-making purposes, each heterogeneous beam type (determined by the two antenna types in each baseline) is treated as a separate mosaicked dataset. We follow a mosaicking and deconvolution algorithm similar to the one presented by \cite{Gueth1995ADeconvolution}, which we briefly summarize. If each observed field produces a map, $\textbf{F}(x,y)$, attenuated by a primary beam, $\textbf{B}(x,y)$, then in order to recover the true sky signal we must correct for the primary beam by dividing each observed field by its corresponding primary beam. We drop the explicit dependence on $x$ and $y$ below for brevity of notation, but note Eqns. \ref{eq:Jxy} and \ref{eq:Nxy} apply as a function of position. To combine the fields into a mosaicked map in units of Jy beam$^{-1}$, we find the weighted mean of the observed fields,
\begin{equation}
  \textbf{J} = \frac{\sum\limits_{i} \textbf{B}_i^2/\sigma_{i}^{2} \times \textbf{F}_i/\textbf{B}_i}{\sum\limits_{i} \textbf{B}_i^2/\sigma_{i}^{2}} = \frac{\sum\limits_{i} \textbf{B}_i/\sigma_{i}^{2} \times \textbf{F}_i}{\sum\limits_{i} \textbf{B}_i^2/\sigma_{i}^{2}},
  \label{eq:Jxy}
\end{equation}
where the subscript $i$ represents each separate mosaic field, and $\sigma_i$ is the noise level in each observed field, determined from the variance of the interferometric visibilities making up each observed field. Each map is truncated where the primary beam, $\textbf{B}_i$, drops below the 10\% level. The noise level in the combined $\textbf{J}$ map is given by,
\begin{equation}
  \textbf{N} = \frac{1} {\sqrt[]{\sum_{i} (\textbf{B}^2_{i} / \sigma_{i}^{2})}}
  \label{eq:Nxy}
\end{equation}
and an SNR map is defined as $\textbf{J} / \textbf{N}$. While the noise is not constant in the resulting $\textbf{J}$ map, the gain of an observed point source at any position in the map is now uniform. A map of the noise level, $\textbf{N}$, is shown in Figure \ref{fig:ms0735_noise}.

The mosaic CLEAN algorithm described by \cite{Gueth1995ADeconvolution} can be summarized as follows. The sky signal is assumed to be made up of a sum of $\delta$-functions, and the measured, or dirty, map is the convolution of the synthesized, or dirty, beam of the telescope with the sky. The dirty map will be contaminated by the sidelobes of the synthesized beam, an artifact of the discreet sampling of $uv$-space by the interferometer. The CLEAN algorithm attempts to fill in the empty spacings by replacing the synthesized beam with a CLEAN beam approximated by a Gaussian fit to the main lobe of the synthesized beam for each mosaicked dataset. A CLEAN map is derived by iteratively finding the peak of the combined SNR map, removing its contribution, and recalculating the $\textbf{J}$ map to form a new SNR map until a SNR cutoff for the peak of the map is reached. The $\delta$-functions at the locations of the peaks are convolved with the CLEAN, or restoring, beam, and the residual map is added back in to preserve the noise in this final CLEAN map. Performing the CLEAN procedure on a SNR map as described above accounts for the varying noise across the map due to the primary beams and reduces the risk of using a noise peak near the edge of a primary beam in $\textbf{J}$ as one of the components of the CLEAN algorithm.

\begin{figure}[h]
  \centering
  \includegraphics[width=1.0\linewidth]{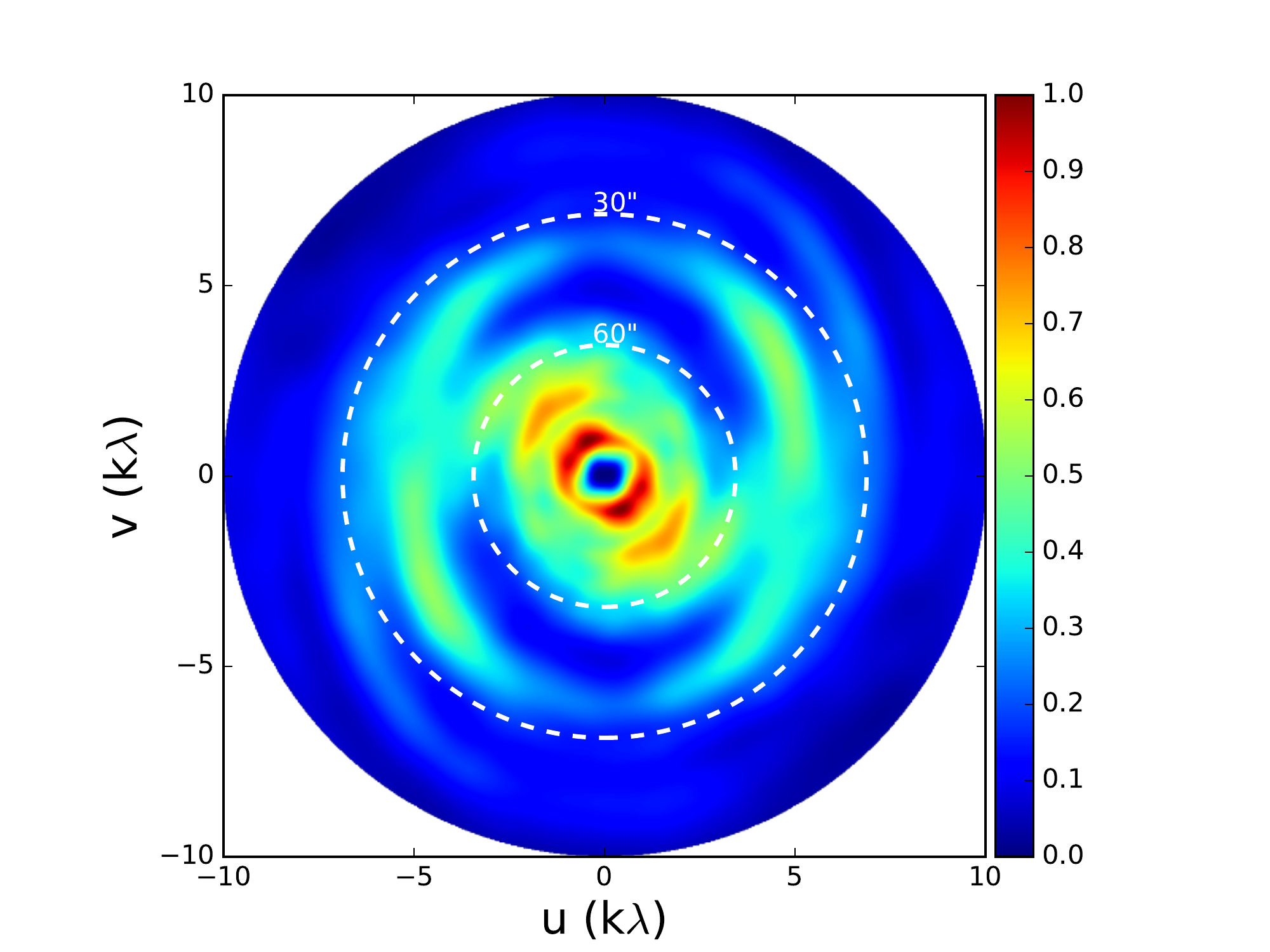}
  \caption{The normalized data weight distribution in the $uv$-plane for CARMA data described in Table \ref{table:observations}. The weights are calculated from the inverse of the variance on each visibility and the extent of each visibility weight is determined by the corresponding primary beam within the heterogeneous array. A small region at $<0.35 \mathrm{k}\lambda$ is inaccessible by CARMA owing to the shadowing limit of the 3.5 m dishes. White, dotted circles corresponding to baselines with $\lambda/B=30\arcsec$ and $60\arcsec$ are shown. The weights are well-matched to the signal expected from the cluster ($\sim 1\arcmin-3\arcmin$) and the cavities ($\sim 30\arcsec-60\arcsec$).}
  \label{fig:ms0735_uvweight}
\end{figure}

In this work, we employ a CLEAN gain of $10\%$ and flux cutoff of $1\sigma$ within a box $6\arcmin$ on a side centered on the central AGN. The resolution of the combined map remains ill-defined because there is a separate restoring beam for each beam-type and mosaicked pointing in the CLEAN maps. An image of the data weight distribution in the $uv$-plane for the data sets described in Table \ref{table:observations} is shown in Figure \ref{fig:ms0735_uvweight}, which provides a more complete picture of the sensitivity of our dataset to different scales. 

\section{Analysis} \label{sec:anal}

To determine the SZ signal associated with the X-ray cavities we first build a model for the rest of the observed cluster. This includes investigating possible extended emission from the radio lobes, and accounting for the central radio source associate with the AGN, any other nearby radio sources, and the SZ contribution from the ICM within which the cavities are embedded, which we refer to as the global ICM. The subsections below detail our modeling for each of these components.

\subsection{Radio Lobe Emission} \label{sec:contamination}

Extended radio lobe emission could contaminate measurements of the SZ effect in these cavities. While cavities are frequently associated with low frequency radio lobes, many clusters do exhibit cavities without detectable high-frequency radio emission due to energy losses from adiabatic expansion of the cavities along with the ongoing inverse-Compton and synchrotron energy losses. An X-ray cavity absent of high frequency radio emission is called a ``ghost cavity'' or ``radio ghost'' \citep{Ensslin1999RadioGhosts}, indicating they are relics of older outbursts, (e.g., \citealt{McNamara2001Discovery2597,Fabian2006AConduction}). Studies of MS0735 at radio frequencies suggest it can also be classified as a radio ghost due to the low break frequency\footnote{The break frequency defines the point above which the spectrum falls steeply.} for the synchrotron emission from the lobes, $\nu_b = 330$ MHz, and most clearly from the lack of observed radio lobes in 8 GHz VLA data \citep{Birzan2008RadiativePower}. 

\begin{figure*}[t]
\centering
  \subfigure[]{
    \includegraphics[width=0.31\linewidth]{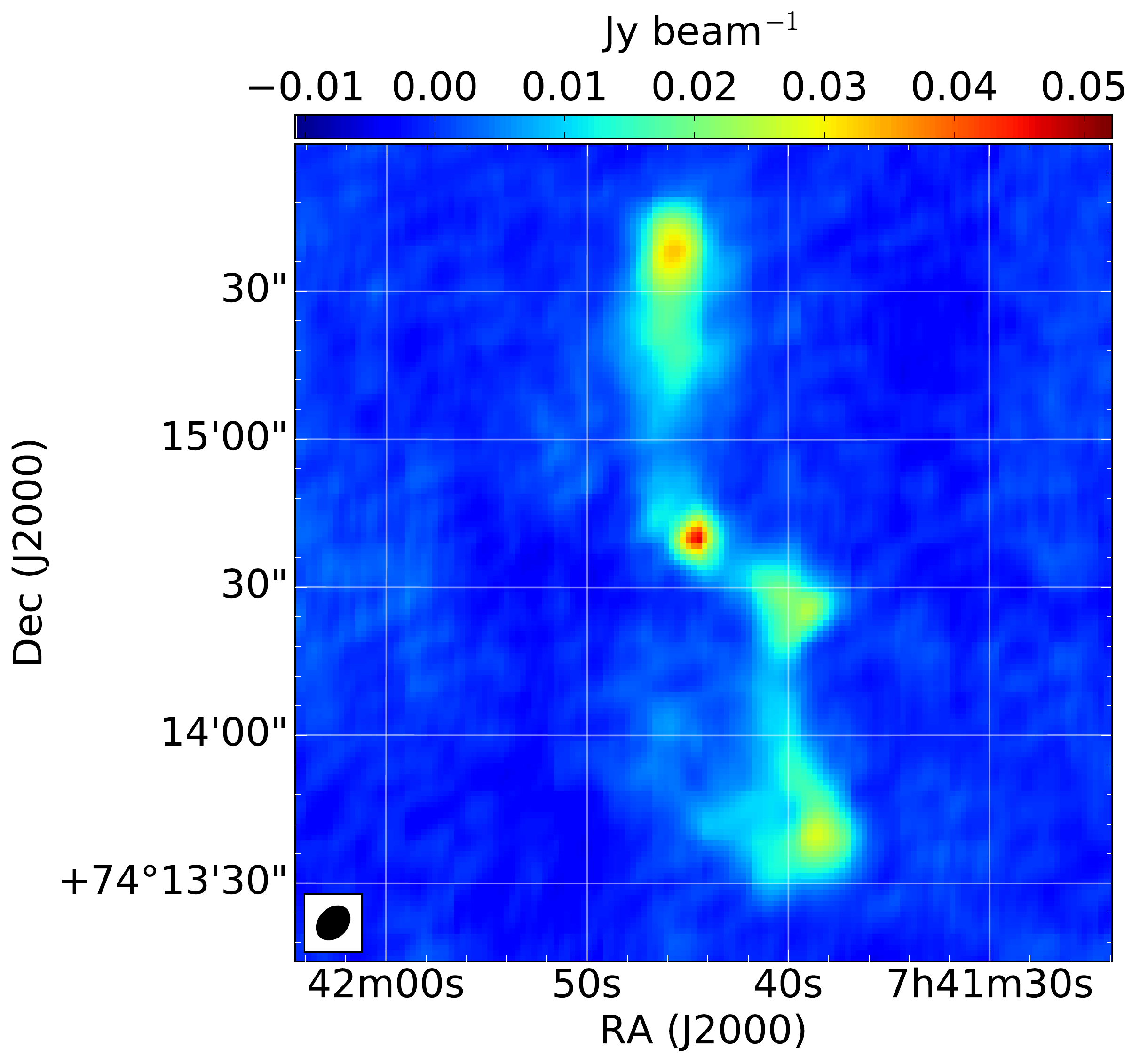}
    \label{fig:ms0735_327}}
  \subfigure[]{
    \includegraphics[width=0.31\linewidth]{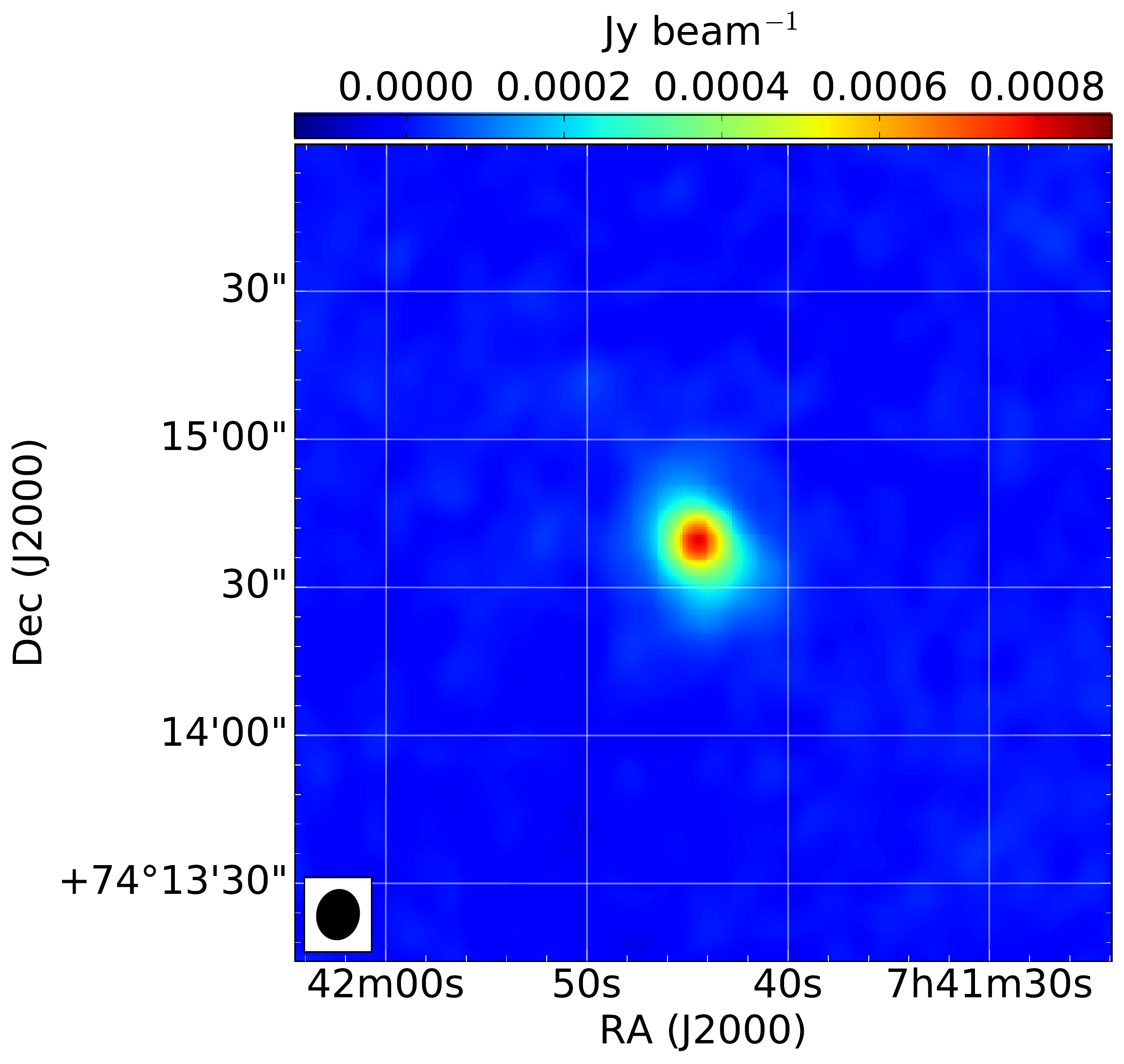}
    \label{fig:ms0735_8}}
  \subfigure[]{
    \includegraphics[width=0.31\linewidth]{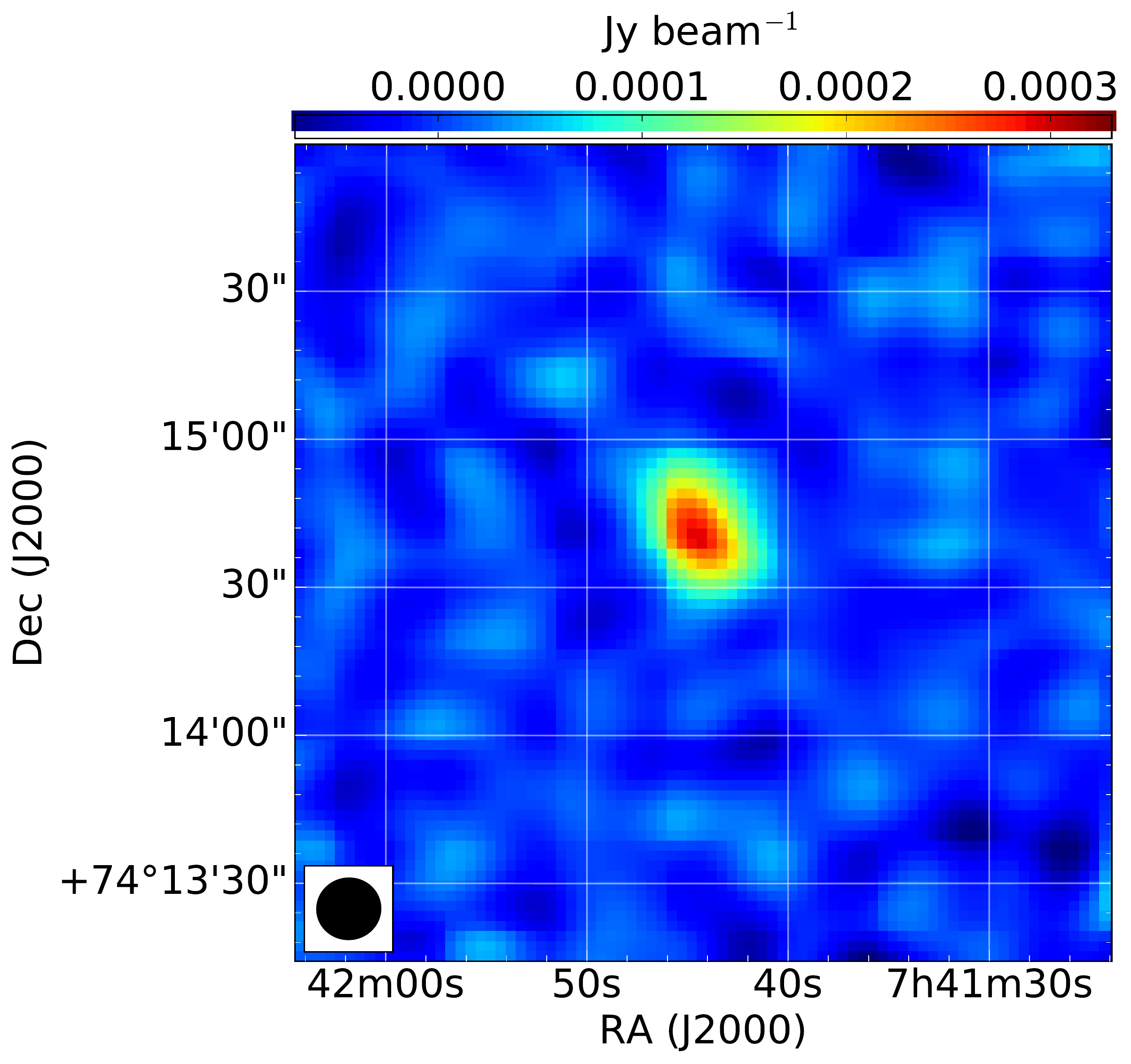}
    \label{fig:ms0735_30}}
  \caption{CLEANed maps of (a) 327 MHz VLA data ($\sigma_\mathrm{rms}=1.1$ mJy beam$^{-1}$), (b) 8.5 GHz VLA data ($\sigma_\mathrm{rms}=0.015$ mJy beam$^{-1}$), and (c) long baseline ($>3 k\lambda$) 30 GHz CARMA data ($\sigma_\mathrm{rms}=0.018$ mJy beam$^{-1}$ at map center) of MS0735. The restoring beam is shown in the bottom left corner of each map. For the CARMA mosaicked map, the smallest restoring beam is shown.}
  \label{fig:central}
\end{figure*}

\begin{deluxetable*}{c c c c c c }[t]
\tablecaption{Central Radio Source Model} \label{tab:central}
\tablehead{Central source & $\Delta x (\arcsec)$ & $\Delta y (\arcsec)$ & major $\sigma (\arcsec)$ & axis ratio & rot. angle ($^\circ$)}
\startdata
Point source & $1.99 \pm 0.1$ & $0.63 \pm 0.1$ & - & - & - \\
Gaussian Halo & $0.58 \pm 0.3$ & $-2.7 \pm 0.3$ & $6.21 \pm 0.3$ & $1.48 \pm 0.1$ & $36.2 \pm 4$ \\
\enddata
\tablecomments{The best-fitting geometry of the two-component model for the central source derived from 8 GHz VLA data (Figure \ref{fig:central}), which is used to model the spatial distribution of the central radio emission in the CARMA data (see section \ref{sec:AGN}). The positional offsets  $\Delta x$ and $\Delta y$ are from the cluster center: $07^\mathrm{h} 41^\mathrm{m} 44^\mathrm{s}, +74^{\circ} 14\arcmin 38\arcsec$. The rotation angle is measure east of north.}
\end{deluxetable*}

To estimate the expected flux density in the lobes at 30 GHz, we use a power-law spectrum, $S \propto \nu^{-\alpha}$, and the reported VLA flux densities in the lobes, $S_{\mathrm{327 MHz}}=0.72$ Jy and $S_{\mathrm{1.4 GHz}}=11.7$ mJy, finding $S_{\mathrm{30 GHz}} \sim 2\ \mu$Jy. The minimum rms noise in our CARMA map is $ \sim 14\ \mu \mathrm{Jy beam}^{-1}$ (Figure \ref{fig:ms0735_noise}), a factor of several above the estimated lobe emission. The estimated $S_{\mathrm{30 GHz}} \sim 2\ \mu$Jy is conservative as the lobes emission is likely distributed across multiple beams, requiring  higher sensitivity to detect, and because the estimate does not account for the observed spectral break, which would make it considerably fainter at 30 GHz (by an additional order of magnitude). We proceed in our analysis under the assumption that the synchrotron emission from the cavities of MS0735 is not a source of significant radio contamination at 30 GHz compared with the noise. VLA observations from \cite{Birzan2008RadiativePower} along with a long baseline ($>3 k\lambda$) CARMA map of MS0735, which is insensitive to the large scale SZ decrement of the cluster, are shown in Figure \ref{fig:central}.

\subsection{Central Radio Source} \label{sec:AGN}

\begin{figure*}[t]
  \centering
  \subfigure{
  \includegraphics[width=0.4\paperwidth]{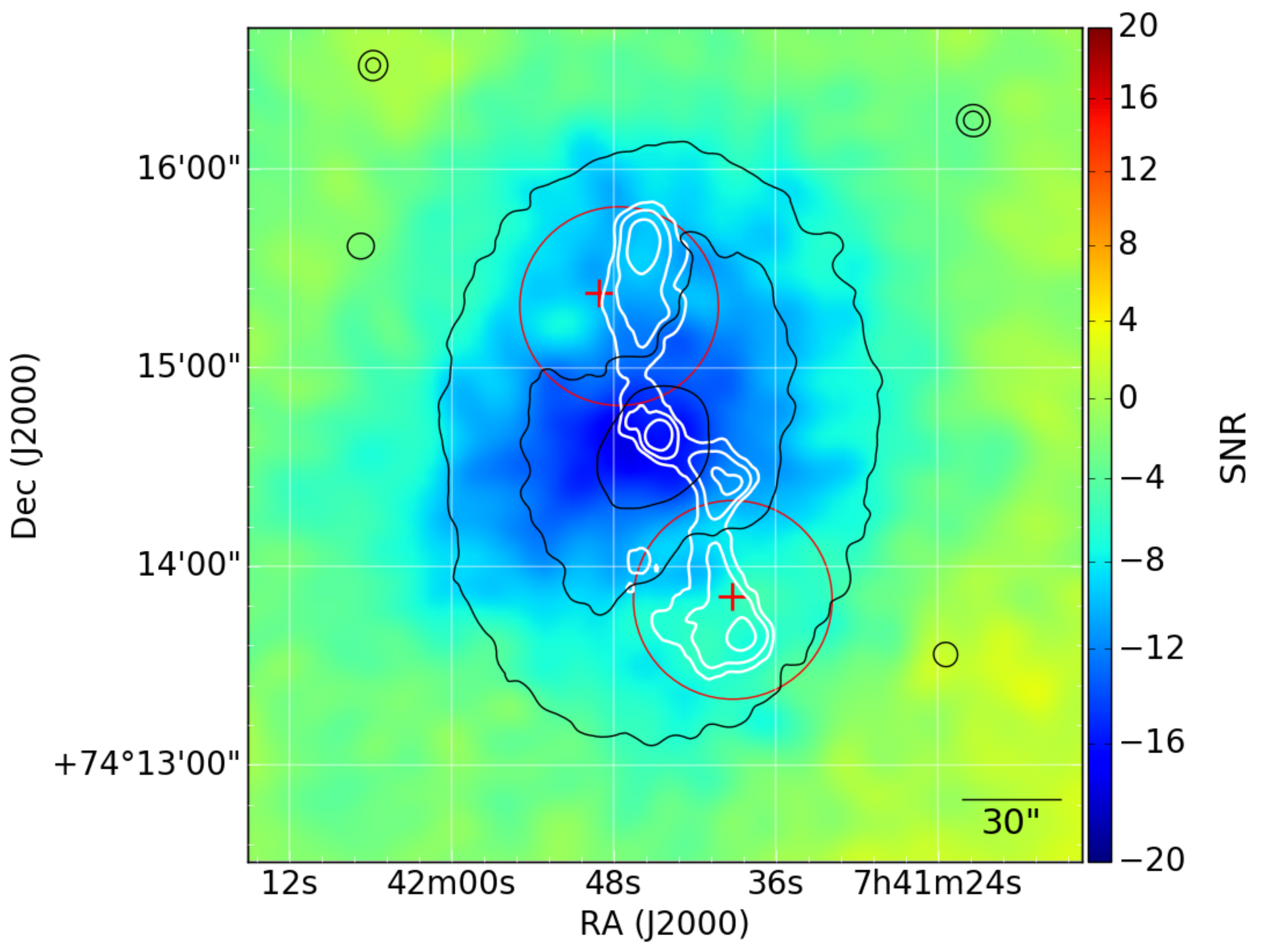}}
  \subfigure{
  \includegraphics[width=0.4\paperwidth]{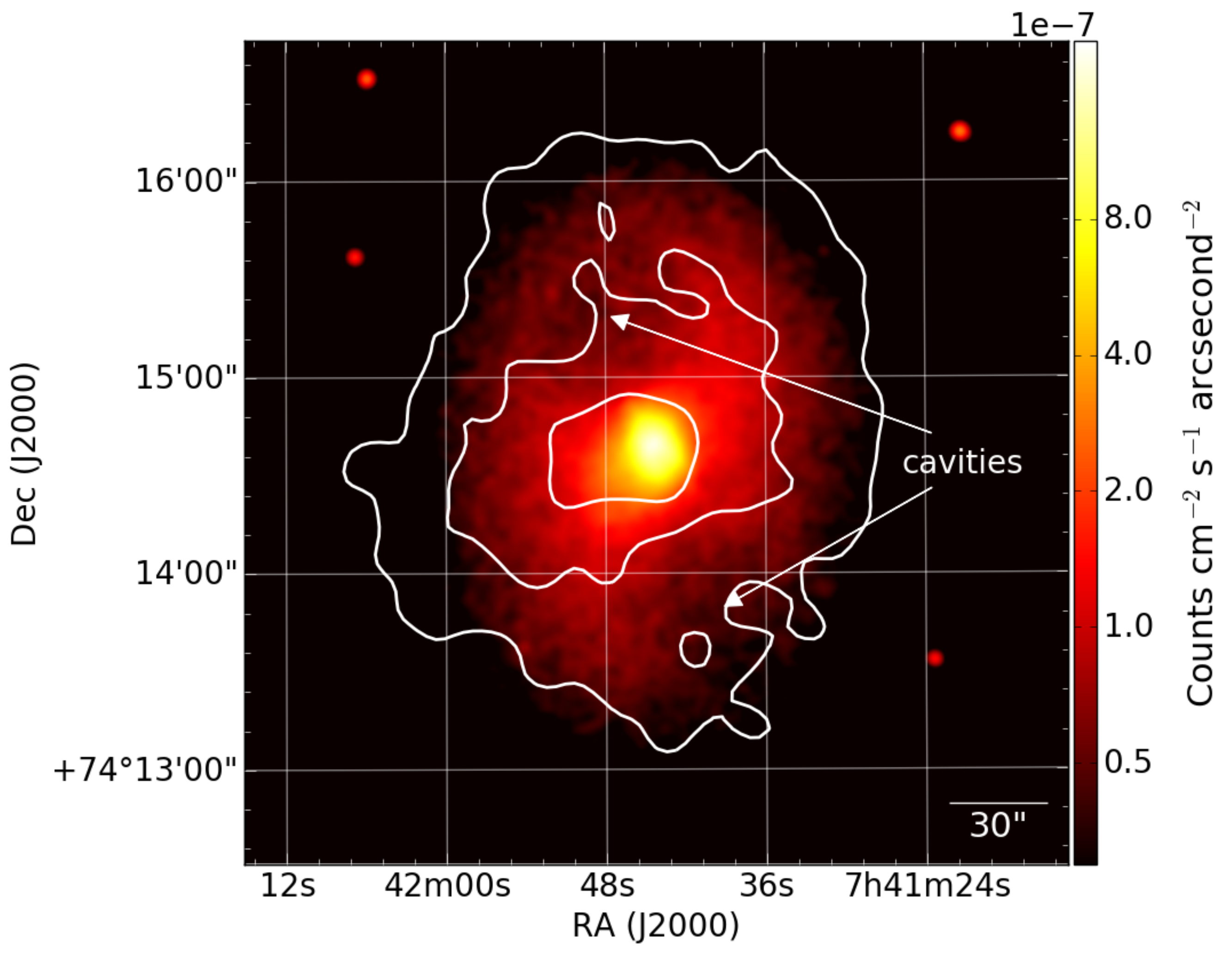}}
  \caption{SZ and X-ray images of MS0735. \textit{On the left}, a CLEAN CARMA map of MS0735 with the central radio source removed (see section \ref{sec:AGN}) representing the total SZ signal in units of SNR. An image of the noise map used to make CARMA SNR maps is shown in Figure \ref{fig:ms0735_noise}. White contours show the 327 MHz VLA observations of the AGN jets at levels of $(0.005,0.01,0.02)$ Jy beam$^{-1}$. Black contours show the Chandra X-ray image (0.5-7 keV) at smoothed levels of $(3.0,7.5,20.0) \times 10^{-8}$ counts cm$^{-2}$ s$^{-1}$ arcsecond$^{-2}$. Red crosses show the centers of the two mosaic pointings of the CARMA-23 observations (see Section \ref{sec:obs}) and red circles show the X-ray identified extent of the cavities. \textit{On the right}, the Chandra X-ray image (0.5-7 keV) Gaussian-smoothed with a 3 pixel kernel radius, with white contours from the CARMA SZ map at levels of $-14\sigma,-10\sigma,-6\sigma$. Note the depressions in both X-ray surface brightness and SZ signal in the regions occupied by the jets.}
  \label{fig:ms0735_image}
\end{figure*}

Low frequency radio observations suggest the central source could be extended enough to be resolved in the CARMA data (Figure \ref{fig:central}). As we are interested in only accounting for this emission in order to remove it, we model the central source heuristically using the 8 GHz VLA data provided by \cite{Birzan2008RadiativePower}. The 8 GHz $uv$-data are modeled with a point source and a surrounding elliptical Gaussian halo of emission. The position, geometry, and extent of this two-component model are fixed to the parameter values found in the 8 GHz data (Table \ref{tab:central}), while the normalization of each component is allowed to vary in the fit to CARMA observations, thus allowing for separate spectral indexes. The central source model is fit simultaneously with cluster models in later analysis to allow for any degeneracy of the extended radio emission with the SZ decrement in the center of the cluster (see section \ref{sec:results} and Figure \ref{fig:pfrommer_triangle}). A CLEAN map of the SZ signal from MS0735 with the central radio source removed (using parameter values from this simultaneous fit) is shown in Figure \ref{fig:ms0735_image}. The total flux density of the central radio source is measured to be $0.35 \pm 0.03$ mJy.

\subsection{Nearby Radio Sources} \label{sec:nearby_source}

There are two point-like 30 GHz sources in the field of view, both located far ($>3$\arcmin) from the cluster center (Table \ref{tab:radio}). The position of each of these sources is consistent with point sources found in 8 GHz VLA observations, and the position of the point source to the south is also consistent with a 1.4 GHz source from the NVSS catalog \citep{Condon1998TheSurvey}. These point sources are modeled simultaneously with the central radio source and SZ cluster models in our analysis.

\begin{deluxetable}{c c c c}[t!]
\tablecaption{Nearby radio sources \label{tab:radio}}
\tablehead{Nearby radio sources & S$_{30 GHz}$ (mJy) & $\Delta x (\arcsec)$ & $\Delta y (\arcsec)$}
\startdata
Southern source & $0.15 \pm 0.04$ & $-2 \pm 3$ & $197 \pm 3$  \\
Northeastern source & $0.19 \pm 0.04$ & $88 \pm 2$ & $151 \pm 2$ \\
\enddata
\tablecomments{List of nearby radio point sources found in the CARMA field of view. The positional offsets $\Delta x$ and $\Delta y$ are from the cluster center: $07^\mathrm{h} 41^\mathrm{m} 44^\mathrm{s}, +74^{\circ} 14\arcmin 38\arcsec$.}
\end{deluxetable}

\subsection{Global ICM}
\label{sec:double-beta}

We next describe an analytical model for the SZ effect we expect to see from a cluster with X-ray cavities. We adapt a model developed for X-ray detectability of these systems \citep{Enlin2002RadioGalaxies} and later proposed for SZ analysis \citep{Pfrommer2005UnveilingEffect}, in which a cavity is embedded in an otherwise smooth ICM. In this model, radio jets have inflated bubbles in the ICM; the approximately spherical bubbles adiabatically expand and quickly settle into near pressure equilibrium with the surrounding medium. In this section, we describe the SZ signal from the extended gas distribution of the smooth ICM, which we refer to as the global ICM, and in the next section, we describe the suppression of the SZ signal from a region occupied by a spherical cavity, which we define as the cavity suppression factor, $f$. 

We use a double $\beta$-model description of the 3-dimensional ICM pressure profile, fit to the \textit{Chandra} X-ray data, as detailed by \cite{Vantyghem2014CyclingClusters}. Specifically, we use the ``combined'' deprojected profile provided by the authors, which is the product of the deprojected density and projected temperature profiles. We additionally require that the ratio of the normalizations of the two $\beta$-model components in the X-ray deprojected pressure profile be preserved in the SZ model. All parameters of this double $\beta$-model are fixed based on the higher resolution X-ray data, except for its overall normalization, which is allowed to vary when fitting the CARMA data.

An overall ellipticity of the cluster gas is apparent in both the X-ray and SZ maps (Figure \ref{fig:ms0735_image}). To best determine the SZ contrast of the cavities, we therefore adopt an elliposoidal model for the global ICM, despite the fact that the X-ray deprojected pressure profile was obtained under the assumption of spherical symmetry. We begin with a triaxial ellipsoidal $\beta$-model for ICM electron pressure,
\begin{equation}
	P_e(x_1,x_2,x_3)=P_{e,0}\bigg(1+\frac{x_1^2}{r_1^2}+\frac{x_2^2}{r_2^2}+\frac{x_3^2}{r_3^2} \bigg)^{-3\beta/2},
    \label{eq:pressure}
\end{equation}
where $r_1$, $r_2$, and $r_3$ are core radii corresponding to each axis. The line-of-sight integral of the pressure gives us the observed Compton-$y$ of the cluster. 

Given the relatively low temperatures measured from X-ray data of MS0735 ($kT < 10$ keV; \citealt{Vantyghem2014CyclingClusters}), we model only the non-relativistic thermal SZ (tSZ) effect of the global ICM \citep{Sunyaev1972TheGalaxies,Birkinshaw1998TheEffect}. The non-relativistic tSZ spectral distortion of the CMB in flux density can be expressed as the product of the thermal Compton-$y$, $y_{th}=\frac{\sigma_\mathrm{T}}{m_e c^2} \int P_{th} dl$, and the spectral shape, $g(x)$, determined from the scattering of CMB photons off a thermal distribution of electrons, where $x=h\nu/kT_\mathrm{CMB}$ is the dimensionless frequency ($x_{30\mathrm{GHz}}=0.5284$). If we integrate the ellipsoidal $\beta$-model along the $x_3$ axis, the Comptonization by the global ICM can be written as \citep{Grego2000The370,Piffaretti2003AsphericalFractions},
\begin{equation}
	y_{cl}(x_1,x_2)=y_{0}\bigg(1+\frac{x_1^2}{r_1^2}+\frac{x_2^2}{r_2^2} \bigg)^{-(3\beta-1)/2}.
\end{equation}
For the ellipsoidal model, we adopt the projected axis ratio (1.37), position angle ($7\deg$ east of north), and cluster center measured from the X-ray image of the outer shock of MS0735 that surrounds the cavities \citep{Vantyghem2014CyclingClusters}. We set the geometric mean of $r_1$ and $r_2$ to the core radius of the corresponding spherical $\beta$-model component of the X-ray derived pressure profile and relate $r_1$ and $r_2$ using the projected axis ratio. A summary of the derived parameters is included in Table \ref{tab:profile}.

\begin{deluxetable}{c c c c c c}[h]
\tablecaption{Pressure Model for Global ICM}
\tablehead{MS0735 & $P_{e,0}$ (keV cm$^{-3})$ & $\beta$ & $r_{1}$ (kpc) & $r_{2}$ (kpc)}
\startdata
Inner $\beta$-model & $0.282$ & $8.93$ & $122$ & $167$ \\
Outer $\beta$-model & $0.074$ & $0.98$ & $249$ & $341$ \\
\enddata
\label{tab:profile}
\tablecomments{List of the pressure model parameters derived from the X-ray data as described in section \ref{sec:double-beta}.}
\end{deluxetable}

\subsection{Cavity Model} \label{sec:bubble}

We model the cavities as spherical regions embedded in the global ICM, whose contribution to the SZ effect differs from the surrounding medium described in the previous section. The shape of the cavities is approximated from the X-ray surface brightness depressions as spherical and 200 kpc across ($\sim 1'$ in diameter) for simplicity, though the surface brightness depressions do exhibit a slightly elliptical shape in the plane of the sky as determined from X-ray data, with ellipticity $\epsilon=1.02$ for the northeast cavity and $\epsilon=1.20$ for the southeast cavity \citep{Vantyghem2014CyclingClusters}. We discuss the impact of line-of-sight geometry for the cluster and cavity models below and in the next section. The positions of the cavities in our SZ model are set by their X-ray identified positions as well.

If a radio bubble contains a non-thermal power-law distribution of relativistic electrons, they will also contribute to the Comptonization of the CMB \citep{Colafrancesco2003TheGalaxies}. Additionally, for very hot thermal plasma of $> 20-50$ keV potentially supporting the cavities, relativistic corrections to the tSZ become important. A generalized formulation of the SZ effect that can account for both non-thermal electrons or very hot thermal gas potentially supporting the cavities is required. The general form of the SZ effect which we employ in this work is detailed by \cite{Ensslin2000ComptonizationPlasma} and \cite{Colafrancesco2003TheGalaxies}, and is briefly described below.

For the general case, we consider the Compton scattering by electrons with density, $n_e$, in a cavity with optical depth,
\begin{equation}
  \tau_{cav} = \sigma_T \int_{cav} n_{e} dl,
\end{equation}
where the subscript \textit{cav} represents the spherical cavity, for which the physical integration limits are determined by the surface of the cavity along the line of sight and are a function of position \citep{Pfrommer2005UnveilingEffect}. The flux scattered to other frequencies from $x$ is $i(x)\tau_{cav}$, where $i(x)$ is the Planckian distribution of the CMB. The flux scattered from other frequencies to $x$ is $j(x)\tau_{cav}$, where
\begin{equation}
  j(x) = \int^\infty_0 dp \int^\infty_0 dt K(t;p) f_e(p) i(x/t)
  \label{eq:j_x}
\end{equation}
is governed by the photon redistribution function for a mono-energetic electron distribution, $K(t;p)$, and a given electron momentum spectrum, $f_e(p)dp$, where $p$ is the normalized electron momentum,\footnote{The normalized electron momentum is $p=\beta \gamma$, where $\beta=v/c=p^2/\sqrt[]{1+p^2}$, and $\gamma=E_e/m_e c^2$. For relativistic electrons, $v \sim c$ and $p \sim \gamma$.} and $t$ is the factor by which the original photon frequency is shifted. The momentum spectrum is normalized so the integral over momentum space is unity. In this work, we use the analytical expression of the redistribution function, $K(t;p)$, for a single Compton scattering derived by \cite{Ensslin2000ComptonizationPlasma}.

The resulting change to the flux density is then given by $\delta i(x) = [j(x)-i(x)]\tau_{cav}$. This result can be more conveniently expressed in terms of a Compton-$y$, $y_{cav} \propto \int_{cav} P_e dl$, and a spectrum, $\tilde{g}(x)$, analogous to the non-relativistic tSZ formulation with the following substitutions (e.g., \citealt{Ensslin2000ComptonizationPlasma,Pfrommer2005UnveilingEffect}):
\begin{equation}
	\delta i(x)_{cav} = [j(x)-i(x)]\tau_{cav}=y_{cav}\tilde{g}(x),
\end{equation}
where
\begin{equation}
	y_{cav} = \frac{\sigma_T}{m_e c^2}\int_{cav} P_e dl,
    \label{eq:y_cav}
\end{equation}
\begin{equation}
	\tilde{g}(x) = [j(x)-i(x)] \frac{m_e c^2}{\langle k\tilde{T}_e \rangle},
    \label{eq:g_x}
\end{equation}
\begin{equation}
	k\tilde{T}_e = P_e/n_e,
\end{equation}
\begin{equation}
    \langle k\tilde{T}_e \rangle=\frac{\int n_e k\tilde{T}_e dl}{\int n_e dl},
\end{equation}
and where $k\tilde{T}_e = P_e/n_e$ is the pseudo-temperature of the particles, which would be the thermodynamic temperature in the case of a thermal electron population. 

We explore the two distinct scenarios of a purely thermal distribution and non-thermal distribution of particles providing pressure support in the cavities (e.g., \citealt{Ensslin2000ComptonizationPlasma}). A thermal distribution of electron momenta,
\begin{equation}
  f_{e,th}(p) = \frac{\beta_{th}}{K_2(\beta_{th})} p^2 \mathrm{exp}(-\beta_{th}\sqrt[]{1+p^2}),
  \label{eqn:thermal}
\end{equation}
recovers the relativistically correct tSZ formulation, where $K_{\nu}$ is the modified Bessel function \citep{Abramowitz1965HandbookTables}, and $\beta_{th}=m_e c^2/kT_e$. For a non-thermal distribution, a power-law electron momentum spectrum, with an upper and lower momentum cutoff, can be written as
\begin{equation}
  f_{e,non-th}(p; \alpha, p_1, p_2) = \frac{(\alpha-1)p^{-\alpha}}{p_1^{1-\alpha}-{p_2^{1-\alpha}}}; p_1 < p < p_2,
  \label{eqn:non_thermal}
\end{equation}
where $\alpha$ is the spectral index of the synchrotron spectrum. The amount of scattering in the non-thermal case will be determined primarily by the value of the minimum momentum of the power-law distribution, $p_1$, for typical values of $\alpha \approx 2.5$ and $p_2 \gg 1$. For MS0735, $\alpha = 2.48$ for the radio lobe emission as determined by VLA observations \citep{Birzan2008RadiativePower}. The pseudo-temperature for the non-thermal case is  \citep{Ensslin2000ComptonizationPlasma},
\begin{equation}
  k\tilde{T}_e = \int_0^\infty dp \ f_{e} (p) \frac{1}{3} p v(p) m_e c.
  \label{eqn:psuedo_temp}
\end{equation}

Geometrically, the cluster model is represented by spherical cavities embedded in the double $\beta$-model. We assume the center of the cavities and center of the cluster lie in the plane of the sky. If we also assume that the shape of the pressure profile (as a function of elliptical radii) within a cavity is the same as outside the cavity (so that the Compton-$y$ in the spherical cavity from Eq. \ref{eq:y_cav} can be obtained from a spherical portion of the pressure profile from Eq. \ref{eq:pressure}), we can express the relative change in flux density from our model as the non-relativistic tSZ contribution from the extended cluster with the spherical cavity removed plus any SZ contribution from a general particle population in a spherical cavity,
\begin{equation}
\label{eqn:y_total}
\delta i (x) = [y_{cl} - y_{cav}] g(x)
	\\+ y_{cav}\tilde{g}(x).
\end{equation}
Here the Comptonization from the double $\beta$-model component is $y_{cl}$ and the Comptonization from the spherical portion of the double $\beta$-model occupied by the cavity is $y_{cav}$ (a slightly altered definition of the $y_{b}$ used in in \citealt{Pfrommer2005UnveilingEffect}). We can then factor $g(x)$ out from Eq. \ref{eqn:y_total} and write the spectral distortion in terms of a ``suppression factor'', $f=1-\tilde{g}(x)/g(x)$, as
\begin{equation}
\label{eqn:y_tot}
\delta i (x) = (y_{cl} - f y_{cav}) g(x).
\end{equation}
We derive a single suppression factor in our model at 30 GHz for simplicity, but note that $\tilde{g}(x)/g(x)$ is very nearly flat across the relatively narrow frequency band of our observations for all cases considered ($\sim 5 \%$ change across the 26-35 GHz band for the most extreme case).

The geometry of these observations -- a projected view of a three-dimensional cluster with three-dimensional cavities inside it -- requires that we impose some assumptions about the line-of-sight structure of the cluster along with the size and shape of the cavities in order to infer an SZ suppression factor within the cavities. In particular, we must assign a line-of-sight core radius ($r_3$ in Eq. \ref{eq:pressure}). Setting $r_3$ to either the major or minor axes of the cluster in the plane of the sky ($r_1$ or $r_2$) derived in section \ref{sec:double-beta} (Table \ref{tab:profile}), provides an exploratory range of suppression factors.

\begin{figure}[t]
  \centering
  \includegraphics[width=0.9\linewidth]{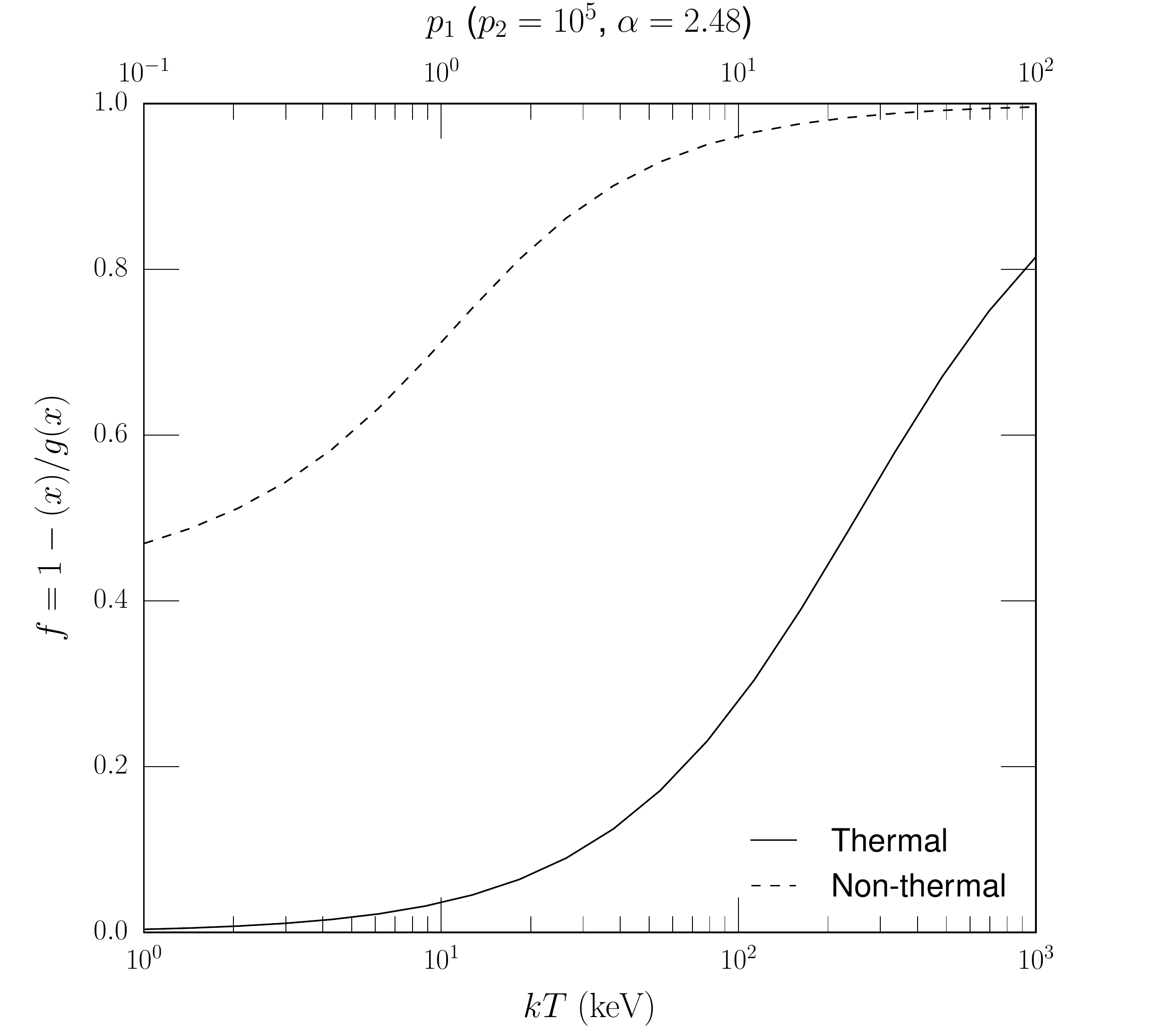}
  \caption{The cavity suppression factor, $f=1-\tilde{g}(x)/g(x)$, at $x_{30\mathrm{GHz}}=0.5284$ as a function of temperature in the case of a thermal (Eq. \ref{eqn:thermal}) electron distribution and minimum momentum cutoff in the case of a non-thermal (Eq. \ref{eqn:non_thermal}) electron distribution. For the non-thermal distribution, $p_2=10^5$ and $\alpha=2.48$. To interpret this value in the context of the cavities we assume that the pressure profile across the cavity follows the elliptical beta-model for the surrounding ICM. A suppression factor of $f=1$ indicates a cavity with no contribution to the SZ signal. Temperature of $>100$ keV are required to suppress the SZ signal by tens of percent in the thermal case, whereas for the non-thermal power-law electron momentum distribution the SZ effect is always suppressed at this level.}
  \label{fig:cavity_suppression_factor}
\end{figure}

The cavity suppression factor is derived from the observed SZ effect from the cluster and the geometry and SZ profile of our model, and so all the details of the cavity composition are contained in the suppression factor. To interpret this value in the context of the cavities we assume that the pressure profile across the cavity follows the elliptical beta-model of the surrounding ICM to constrain the pressure throughout the cavity. The suppression factor fit to the data can then be tied back to a thermal temperature if we assume a single electron temperature throughout the cavity, by use of the thermal electron momentum distribution (Eq. \ref{eqn:thermal}) to obtain the scattered spectrum (Eq. \ref{eq:j_x}), and then to derive the SZ spectrum, $\tilde{g}(x)$ (Eq. \ref{eq:g_x}) and the resulting suppression factor, $f=1-\tilde{g}(x)/g(x)$ (Figure \ref{fig:cavity_suppression_factor}). For the non-thermal case, the suppression factor fit to the data can be tied back to a minimum momentum for a power-law spectrum if we assume a spectral index, an upper momentum cutoff, and a single electron momentum distribution throughout the cavity. Here, the equation for the power-law momentum spectrum (Eq. \ref{eqn:non_thermal}) is used to obtain the pseudo-temperature (Eq. \ref{eqn:psuedo_temp}) and the scattered spectrum (Eq. \ref{eq:j_x}) to find the non-thermal SZ spectrum (Eq. \ref{eq:g_x}) and the resulting suppression factor, $f=1-\tilde{g}(x)/g(x)$ (Figure \ref{fig:cavity_suppression_factor}). 

If the suppression factor is $\sim 1$ then there is minimal SZ contribution from the cavity, which implies the pressure support in the cavity is predominantly by a non-thermal electron distribution, a thermal distribution of very hot gas, $kT_e >1000$ keV, or magnetic fields (which do not contribute to the SZ effect). A suppression factor of $\sim 0$ is consistent with thermal gas of ambient temperature providing the pressure support in the cavities. The equivalent suppression factors corresponding to a few distinct scenarios are shown with the results of our analysis in Figure \ref{fig:cavity_histogram}. In practice, the observed suppression may reflect pressure support from more than one of these distinct scenarios.

\section{Results} \label{sec:results}

In this section, we fit the analytical model components described above to the CARMA data using an MCMC routine. We begin with special case of $f=0$, in which there is no suppression of the SZ effect associated with the cavities, and continue with the inclusion of the cavity model, allowing the cavity suppression factor, $f$, to be fit along with the other model components. We image the residuals resulting from removing the best-fit model components from the CARMA data to reveal the signal from the cavities. Finally, we compare the posterior distributions for the cavity suppression factor to the physical scenarios for pressure support described in section \ref{sec:bubble}.

\begin{figure}[t]
  \centering
  \includegraphics[width=0.9\linewidth]{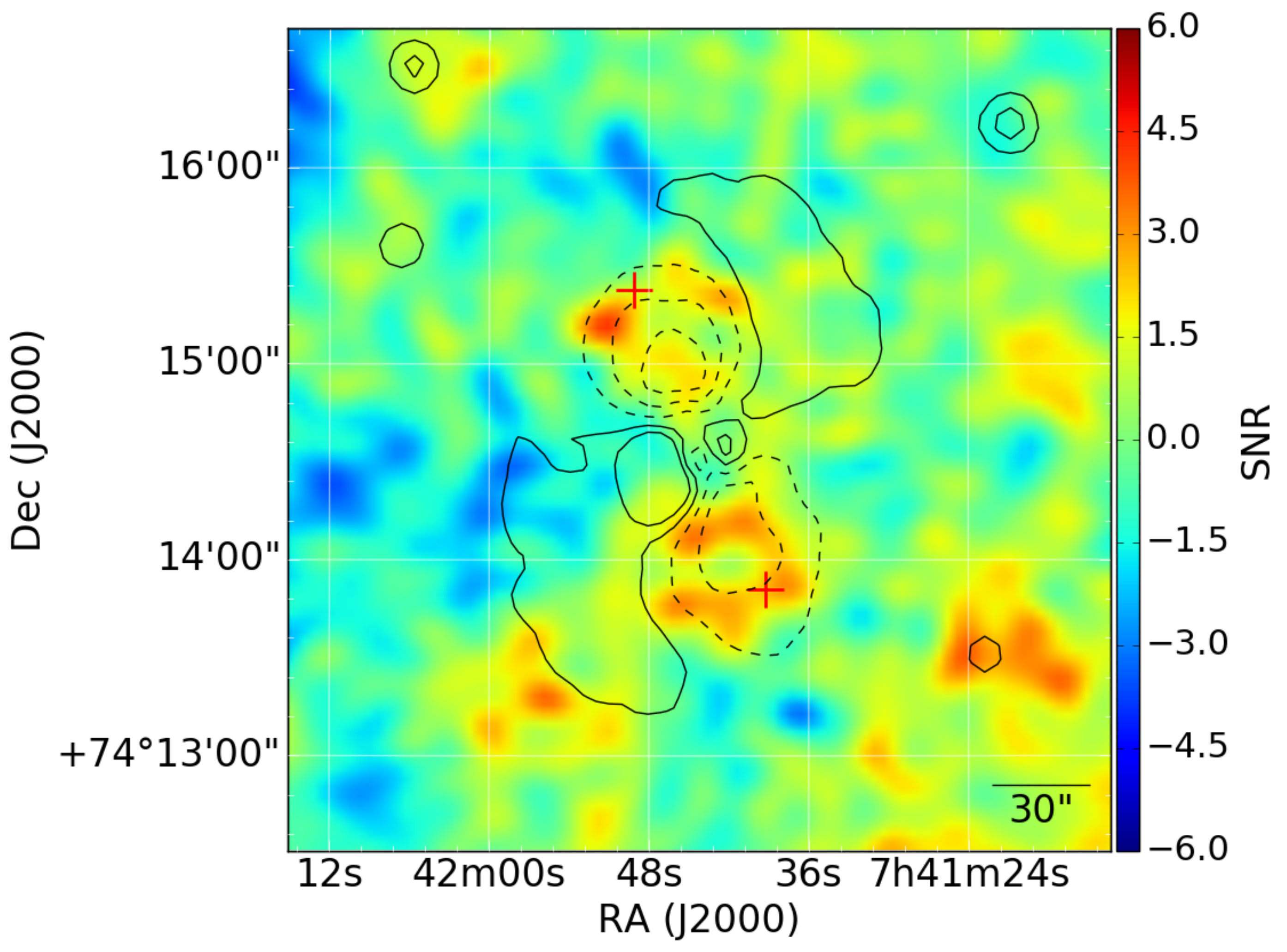}
  \caption{A CLEANed map in units of SNR after removing the best-fit model for the special case $f=0$ from the CARMA data. In this model, only the X-ray derived double-$\beta$ model is used to model the SZ effect from the cluster. An image of the noise map used to make SNR maps is shown in Figure \ref{fig:ms0735_noise}. For comparison, black contours show a smoothed \textit{Chandra} X-ray (0.5-7 keV) map with the best fit double $\beta$-model removed as in Figure 3 of \cite{Vantyghem2014CyclingClusters} at levels of $(-5,-3,-1,1,3) \times 10^{-8}$ counts cm$^{-2}$ s$^{-1}$ arcsecond$^{-2}$. Solid and dotted black lines indicate an excess and deficit of X-ray surface brightness, respectively, when compared to the best-fit double $\beta$-model. The positive residuals are coincident with the X-ray cavities, indicating a smooth double $\beta$-model over-predicts the SZ signal in the cavities, which we expect from cavities that are supported by non-thermal or very hot thermal particles (see section \ref{sec:bubble}).}
  \label{fig:double_beta_residuals}
\end{figure}

For the special case of $f=0$, the signal for the cluster is represented by the X-ray derived smooth double $\beta$-model and the central and nearby radio sources. The residual map after removing the best-fit model components is shown in Figure \ref{fig:double_beta_residuals}. We observe positive residuals of up to $\sim 3 \sigma$ per beam in the regions occupied by the X-ray cavities. Positive residuals indicate the smooth double $\beta$-model over-predicts the SZ signal in the cavities, which we expect from bubbles that are either supported by non-thermal plasma or hot thermal plasma in excess of many tens of keV (i.e., $f>0$; see section \ref{sec:bubble}). The residuals show remarkable resemblance to an analogous image of the a best fit double $\beta$-model removed from the X-ray surface brightness, shown in the Figure 3 of \cite{Vantyghem2014CyclingClusters}. We reproduce that image and overlay the contours on Figure \ref{fig:double_beta_residuals} for comparison.

\begin{figure*}[t]
  \centering
  \subfigure{
    \includegraphics[width=0.31\linewidth]{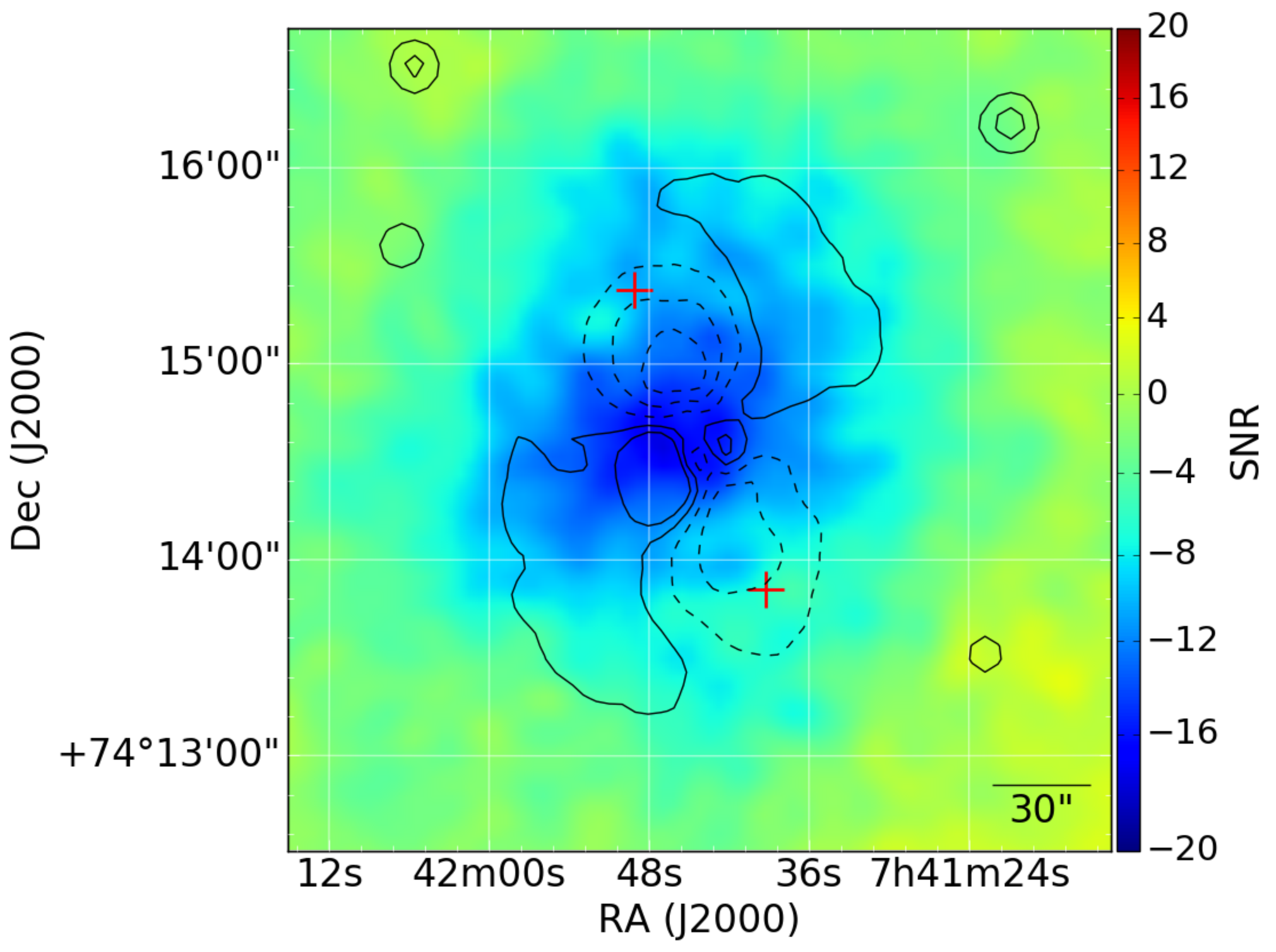}
    \label{fig:ms0735_ptsrc_res}}
  \subfigure{
    \includegraphics[width=0.31\linewidth]{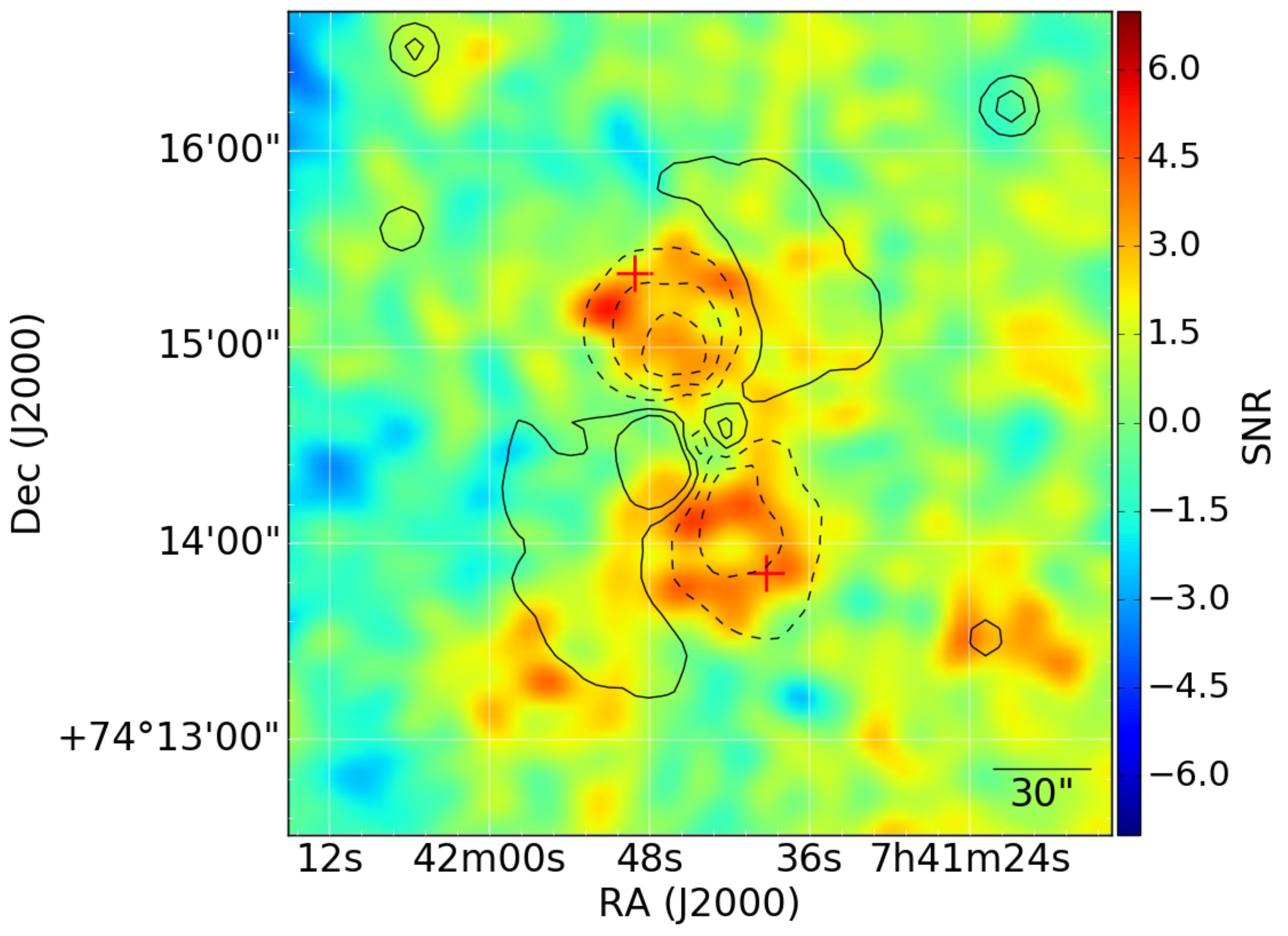}
    \label{fig:ms0735_beta_res}}
  \subfigure{
    \includegraphics[width=0.31\linewidth]{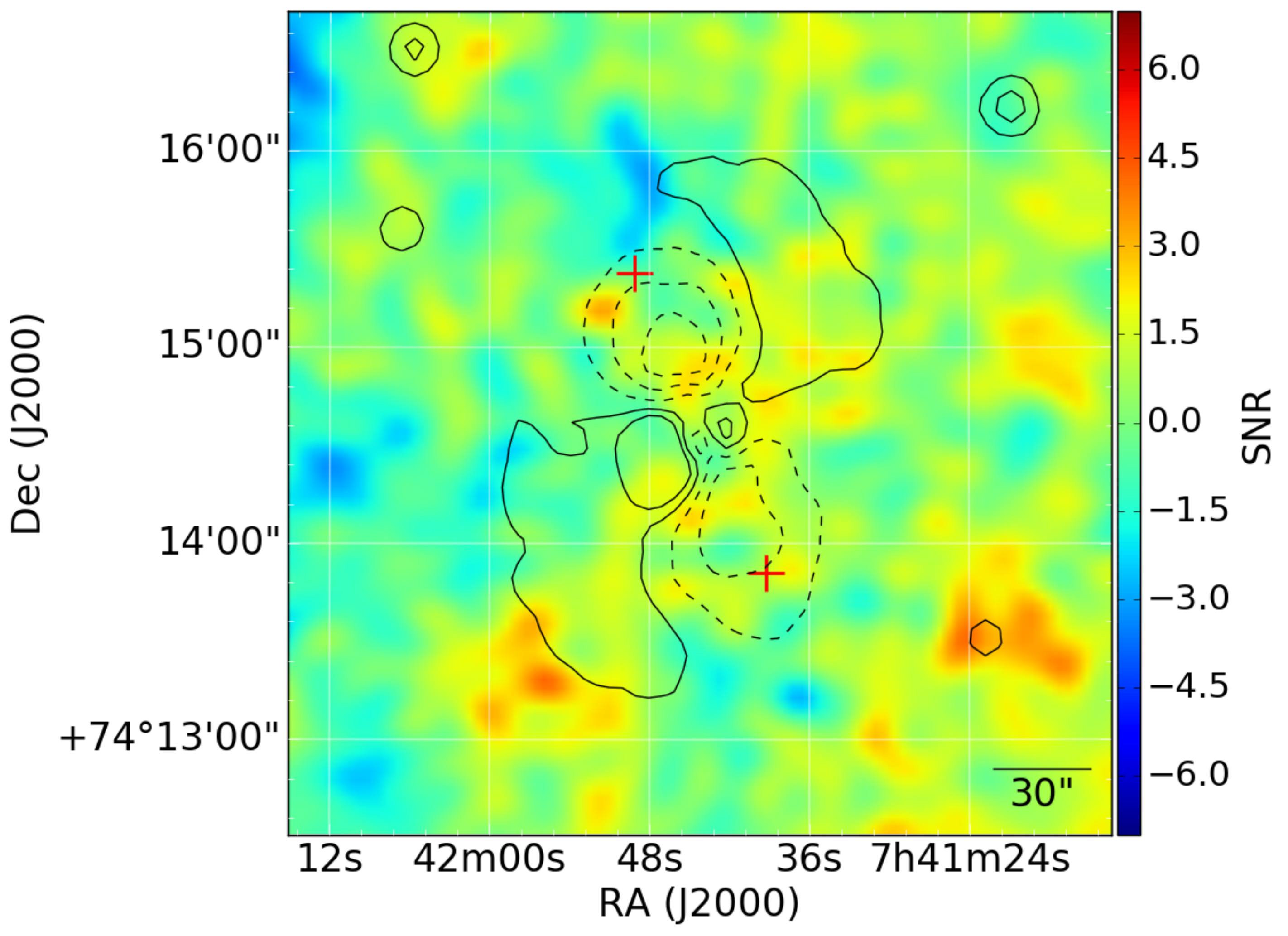}
    \label{fig:ms0735_pfr_res}}
  \subfigure{
    \includegraphics[width=0.31\linewidth]{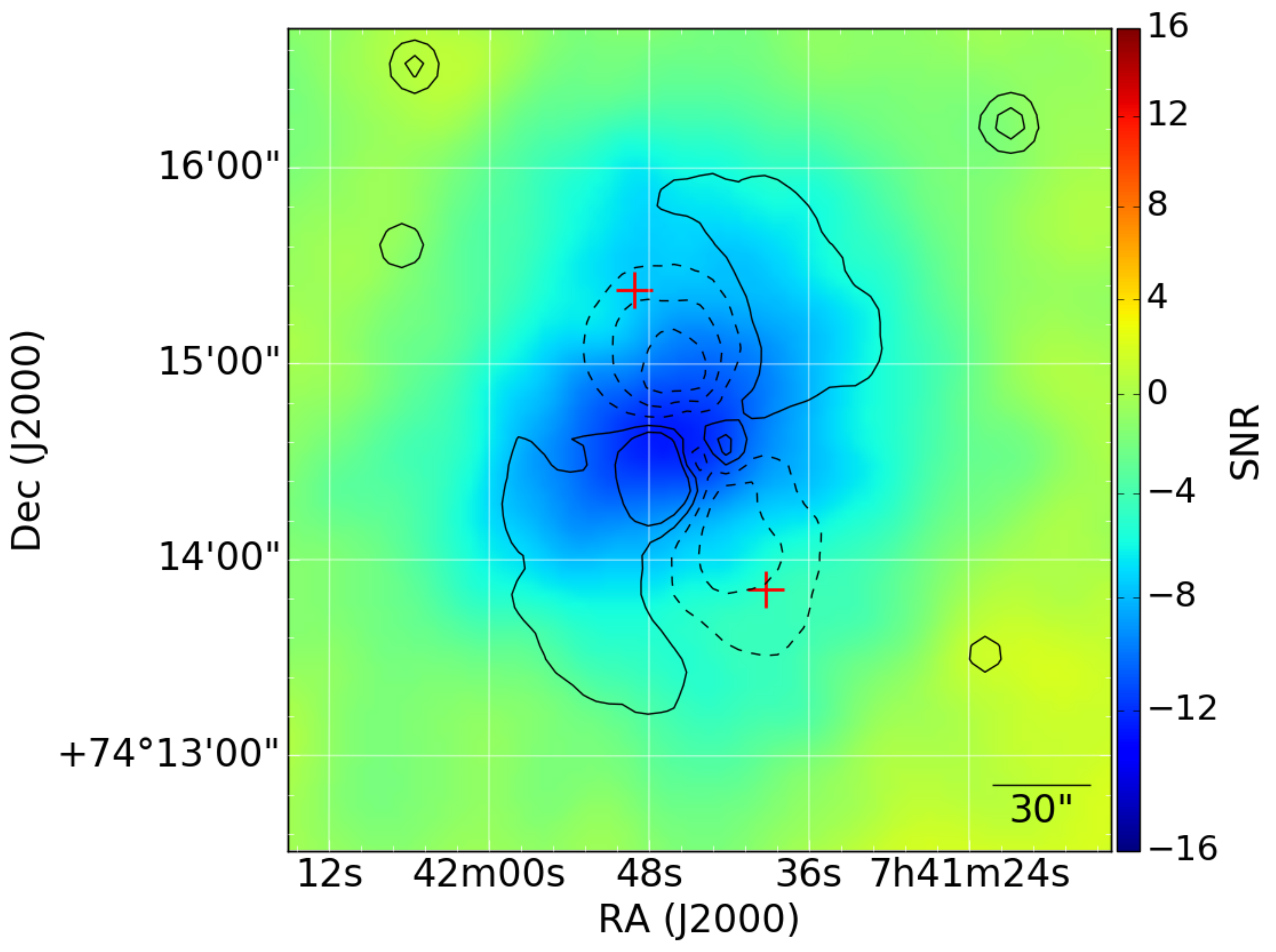}
    \label{fig:ms0735_ptsrc_res_uvtaper}}
  \subfigure{
    \includegraphics[width=0.31\linewidth]{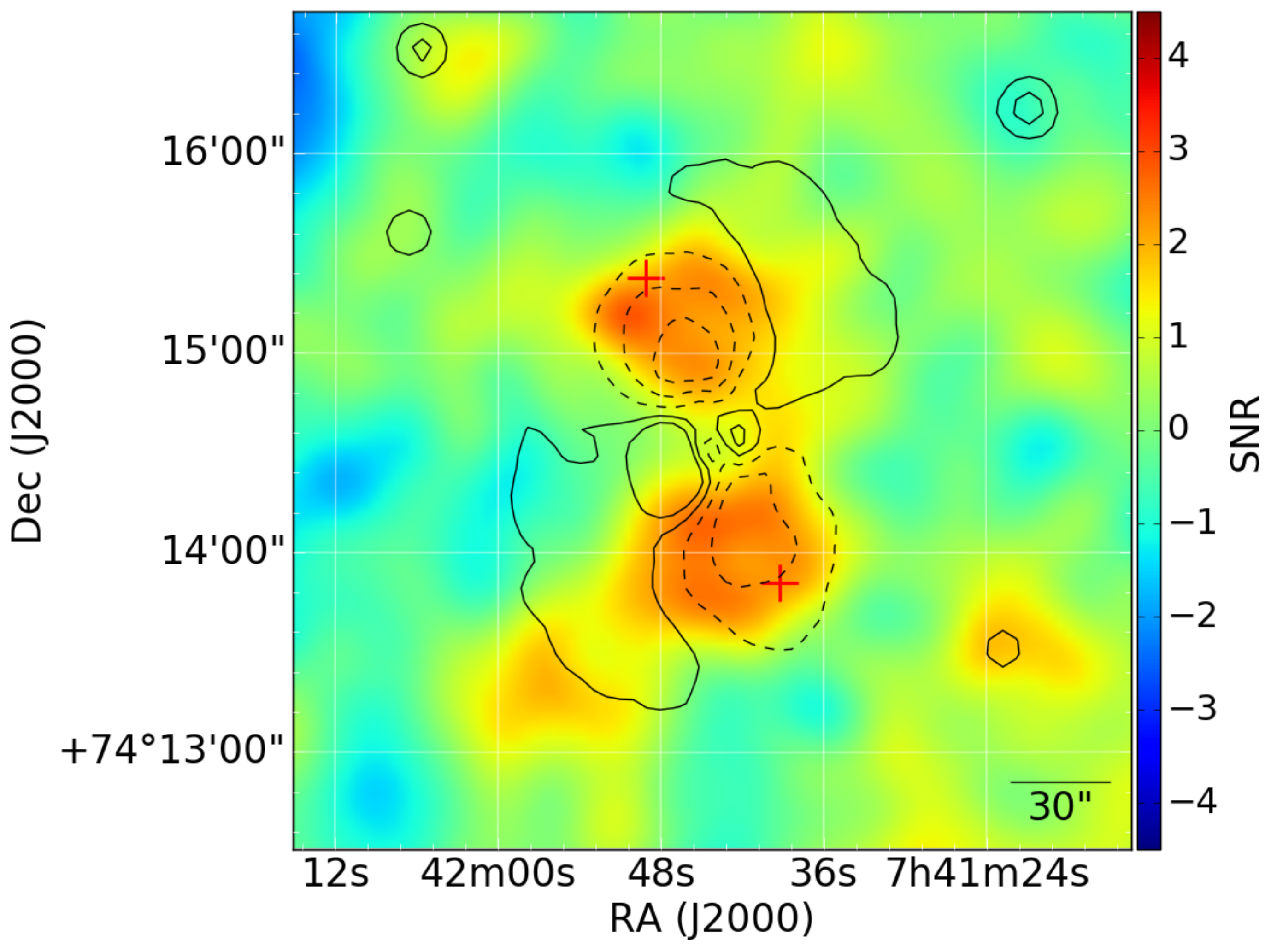}
    \label{fig:ms0735_beta_res_uvtaper}}
  \subfigure{
    \includegraphics[width=0.31\linewidth]{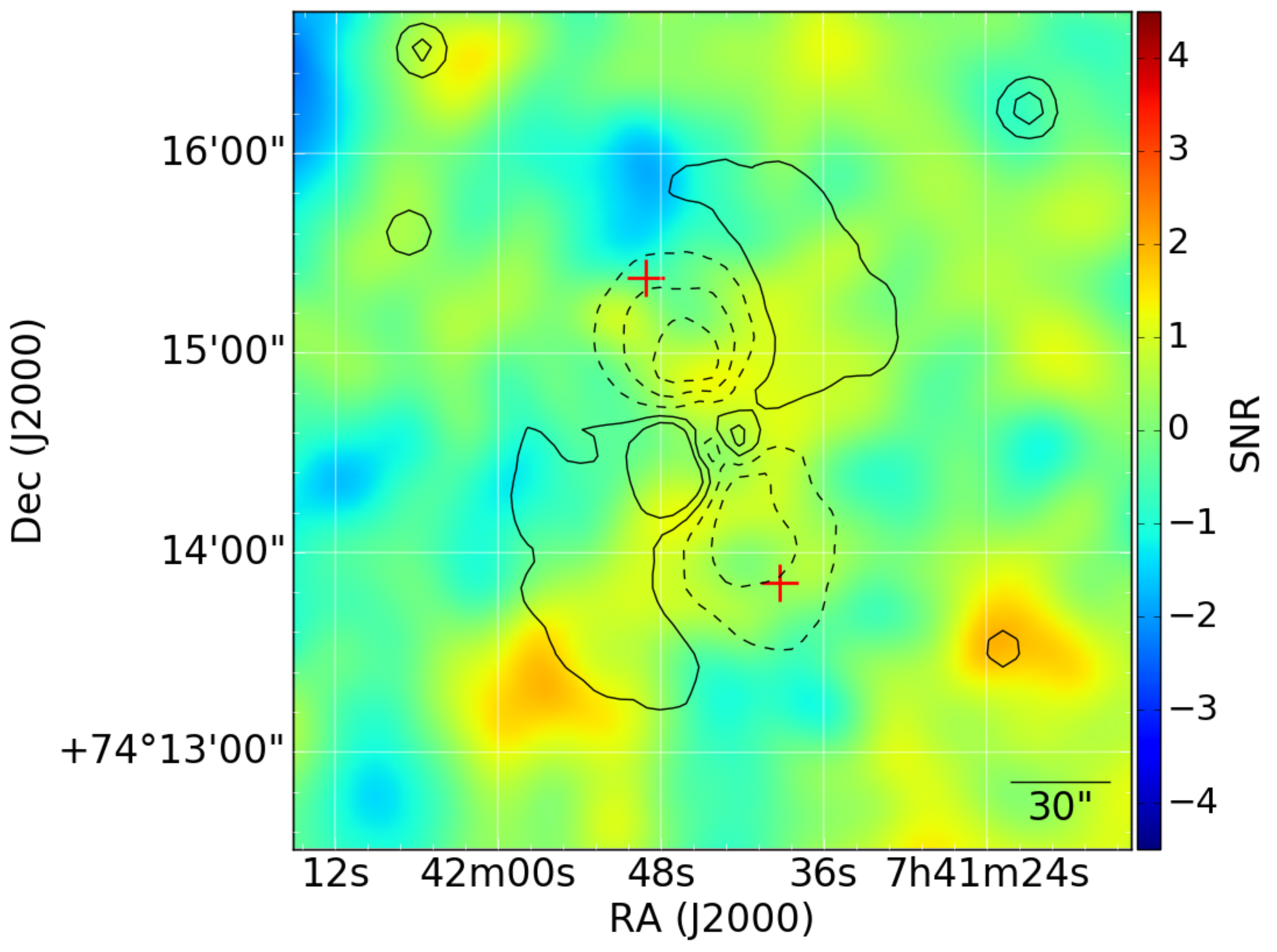}
    \label{fig:ms0735_pfr_res_uvtaper}}
  \caption{CLEANed maps in units of SNR after removing the best-fit model components from the CARMA data. An image of the noise map used to make CARMA SNR maps is shown in Figure \ref{fig:ms0735_noise}. A Gaussian $uv$-taper at $5 \mathrm{k}\lambda$ has been applied to produce the smoothed versions of the maps seen in the bottom panels. \textit{Left:} the central and nearby radio emissive sources are removed to image the total SZ signal (see also the left panel of Figure \ref{fig:ms0735_image}). \textit{Center:} the radio emissive sources and double $\beta$-model are removed, representing our detection of the cavities, and improving on the map shown in Figure \ref{fig:double_beta_residuals}. \textit{Right:} the radio emissive sources, double $\beta$-model, and cavities are removed from the CARMA data, showing our model accounts for nearly all of the observed signal. The colorscale is shown next to each map. For comparison, black contours show the binned \textit{Chandra} X-ray (0.5-7 keV) map with the best-fit double $\beta$-model removed at levels of $(-5,-3,-1,1,3) \times 10^{-8}$ count cm$^{-2}$ s$^{-1}$ arcsecond$^{-2}$ as in Figure 3 of \cite{Vantyghem2014CyclingClusters}. Solid and dotted black lines indicate an excess and deficit of X-ray surface brightness, respectively, when compared to the best-fit double beta-model.}
  \label{fig:pfrommer_residuals}
\end{figure*}

We proceed by including the cavity model by fitting for the cavity suppression factor, $f$. As the expected signal from the cavities is weak, to improve our statistics we fit both cavities with a single suppression factor under the assumption that they have similar composition. The residuals after removing the best-fit model components in this case are shown in Figure \ref{fig:pfrommer_residuals}. The bottom panels of Figure \ref{fig:pfrommer_residuals} shows smoothed versions of the images in the top panels, produced by applying a Gaussian $uv$-taper at $5 k\lambda$. In the left panels, the radio emissive model components are removed, showing the total SZ signal from MS0735 (see also the left panel of Figure \ref{fig:ms0735_image}). In the center panel, both the emissive sources and double $\beta$-model are removed from the CARMA data to reveal the prominent cavities. As the SZ contribution (suppression) from the cavities is now included as a free parameter, the derived double $\beta$-model better represents the contribution to the SZ effect from the global ICM than the special case of $f=0$ shown in Figure \ref{fig:double_beta_residuals}. We now observe positive residual in excess of up to $\sim 5.5 \sigma$ per beam in the regions occupied by the cavities and a better accounting of the SZ effect (smaller negative residuals) from the surrounding ICM. In the rightmost panels of Figure \ref{fig:pfrommer_residuals}, all model components are removed from the CARMA data, demonstrating our model consistently accounts for nearly all of the observed signal.

\begin{figure}[t]
  \centering
  \includegraphics[width=0.9\linewidth]{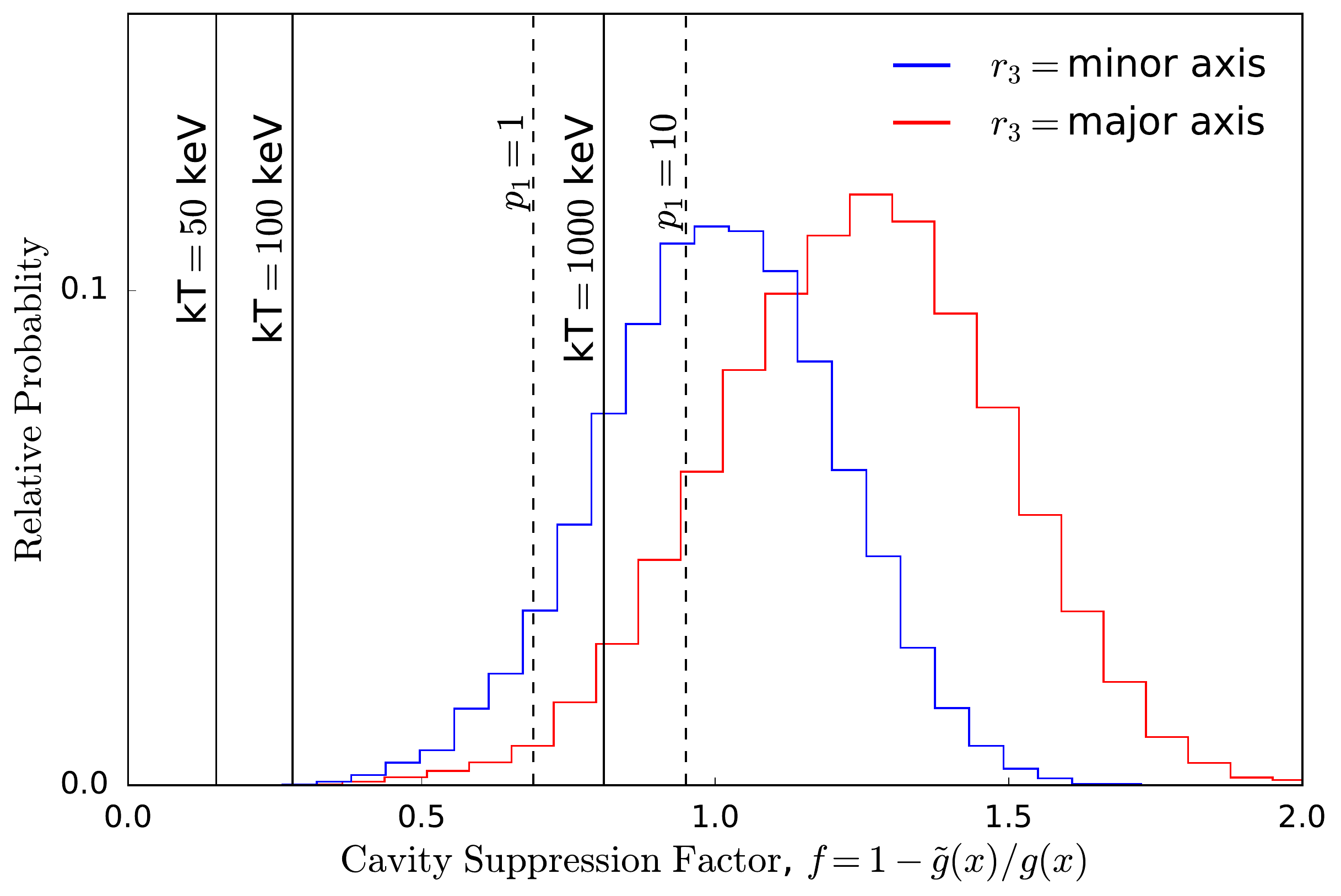}
  \caption{The posterior probability distribution for the cavity suppression factor from an MCMC fit with assumed line-of-sight depth for the cluster of the minor (blue) and major (red) axis of the elliptical cluster model described in \ref{sec:double-beta}. The cavity suppression factor expected from several distinct physical scenarios from the electron distributions described in Eqns. \ref{eqn:thermal} and \ref{eqn:non_thermal} are included as vertical lines. In the non-thermal cases, $\alpha=2.48$ and $p_2=10^5$.}
  \label{fig:cavity_histogram}
\end{figure}

The sharp contrast observed in the center panels of Figure \ref{fig:pfrommer_residuals}, coincident with the X-ray identified cavities, shows a clear detection of the cavities of MS0735 through the SZ effect. Supplying a line-of-sight depth to the model, we constrain the properties of the cavities via the suppression factor described in section \ref{sec:bubble}. The derived suppression factor when adopting the minor and major axes of the elliptical cluster model as the line-of-sight depth of the cluster are $f = 0.98 \pm 0.2$ and $f = 1.31 \pm 0.26$, respectively (Table \ref{tab:results}). The posterior probability distributions for the suppression factor from the MCMC fits, along with values of $f$ corresponding to representative values for the non-thermal and thermal models for pressure support (see section \ref{sec:bubble}) are shown in Figure \ref{fig:cavity_histogram}. 

\begin{figure}[htb]
  \centering
  \includegraphics[width=1.1\linewidth]{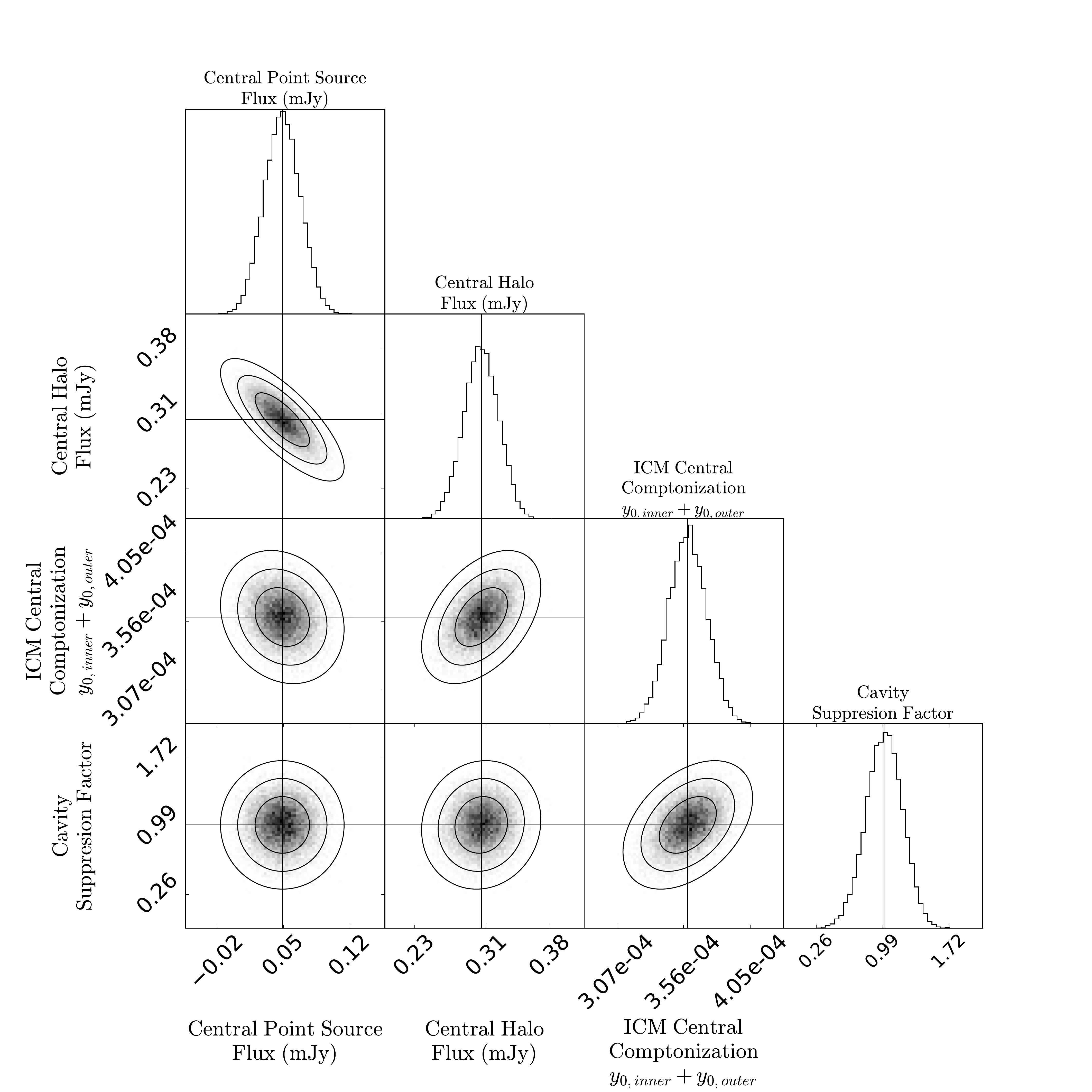}
  \caption{Marginalized constraints on a subset of model parameters describing the emissive radio sources, the global ICM, and cavities. In this model, the minor axis of the projected double $\beta$-model is adopted as the line-of-sight core radius to derive the suppression factor. From left to right, the parameters shown are the flux of point source emission from the central AGN, the flux of the extended emission from the central AGN, the sum of the central Compton-$y$ parameters for the inner and outer components of the double $\beta$-model, and the single cavity suppression factor describing both cavities. The $(1\sigma, 2\sigma, 3\sigma)$ confidence ellipses determined from the covariance of the variable parameters are also shown.}
  \label{fig:pfrommer_triangle}
\end{figure}

Our results disfavor support of the cavities by thermal plasma of less than several hundreds to thousands of keV, whereas pressure support from non-thermal gas or magnetic fields is allowed. Pressure support by thermal plasma of less than $kT_e \sim 150$ keV is excluded at the $99.7\%$ level when assuming a line-of-sight core radius equal to the projected minor-axis core radius of the elliptical cluster model (the more conservative case considered). Marginalized constraints on several model parameters are shown in Figure \ref{fig:pfrommer_triangle}. Results from a fit in which the suppression factors of the two cavities are independent are also included in Table \ref{tab:results}.

%We note here the tentative ``ring''-like structure of the positive residuals observed in the upper-center panel of Figure \ref{fig:pfrommer_residuals}. Though the resolution and SNR do not warrant significant speculation, the shape of the residuals may be indicative of underlying structure within the cavity \citep{Dursi2008DrapingEffects,Braithwaite2010MagnetohydrodynamicMedium,Prokhorov2010ComptonizationCocoons}.%

Suppression factor of $f>1$ are unphysical in our model, but could result simply from statistical noise. Another possibility is that we have assumed an incorrect line-of-sight depth for either the global ICM or the cavities. As we have noted, the southern cavity shows an elliptical geometry in the plane of the sky. A cavity elongated along the line of sight will exhibit higher SZ suppression and would increase the value of $f$, while a cavity compressed along the line of sight would decrease the value of $f$. In our geometric model, a cavity elongated along the line of sight by $20\%$ would correspond to a $\sim 20\%$ increase in $f$. Cavity geometry may also account for differences in measured suppression factors in the model in which they are derived independently for each cavity (see Table \ref{tab:results}).

\begin{deluxetable}{c | c c}[t]
\tablecaption{Measured Cavity Suppression Factors \label{tab:results}} 
\tablehead{Cavity & \multicolumn{2}{c}{Line-of-sight Depth} \\
           Suppression Factor & Minor axis & Major axis}
\startdata
Linked cavities & $0.98 \pm 0.2$  & $1.31 \pm 0.26$ \\
\tableline
North cavity    & $0.81 \pm 0.25$ & $1.01 \pm 0.3$  \\ 
South cavity    & $1.14 \pm 0.32$ & $1.34 \pm 0.4$  \\ 
\enddata
\tablecomments{Measured cavity suppression factors inferred with a given line-of-sight core radius for the cluster (see section \ref{sec:bubble} for details). In the top row, the cavity suppression factors in the north and south cavity are derived using a common (linked) value. In the bottom two rows, the suppression factors of each cavity are derived independent of the other.}
\end{deluxetable}

An F-test comparing the model with $f=0$ ($399230-9$ degrees of freedom, $\chi^2=429397.41$) to the model in which a common suppression factor for the cavities is a free parameter ($399230-10$ degrees of freedom, $\chi^2=429376.13$) yields an F-statistic of 19.79 and associated p-value of $8.6 \times 10^{-6}$ (equivalent to $4.4\sigma$ significance), indicating that the addition of the cavity model significantly improves the fit. This improvement can be seen visually in the residual maps presented in right most panels of Figure \ref{fig:pfrommer_residuals}, compared with those in Figure \ref{fig:double_beta_residuals}. There is no significant improvement in the fit when directly comparing the model where the cavities have a common suppression factor to the model with independent suppression factors for each cavity ($399230-11$ degrees of freedom, $\chi^2=429375.14$).

\section{Discussion \& Conclusion} \label{sec:conclusion}

We have presented high resolution 30 GHz observations of the SZ effect from MS0735, which hosts two giant X-ray cavities. We observe a clear deficit in the SZ signal at the location of the X-ray identified cavities and radio jets associated with the AGN (see center panel of Figure \ref{fig:pfrommer_residuals}). This result represents the first detection of this phenomenon through the SZ effect. Assuming that the cavities are approximately spherical and in approximate pressure balance with the surrounding ICM, we find that the suppression of the SZ effect within the cavities, compared with the surrounding gas, is nearly total. This indicates that if the internal pressure of the cavities is supported by thermal plasma its temperature must be several hundreds to thousands of keV.

Alternatively, the cavities may be supported primarily by non-thermal relativistic particles or magnetic fields (Figure \ref{fig:cavity_histogram}). Adiabatic expansion of the cavity would tend towards non-thermal pressure support owing to the differing adiabatic indexes of cosmic ray (non-thermal and relativistic) protons ($\gamma = 4/3$) and non-relativistic ($kT << m_p c^2$) thermal gas ($\gamma = 5/3$). If even a small fraction of the AGN energy accelerates a population of cosmic ray protons, then as the cavity adiabatically expands, the internal pressure of the cavity will become dominated by the relativistic particles. This physical argument for non-thermal support is, however, weakened if the jet continuously supplies a population of very hot thermal particles throughout the inflation of the cavity or if the cavities do not expand adiabatically. Magnetic fields, which have no SZ signature, may also play an important structural role in the support of cavities \citep{Dursi2008DrapingEffects,Braithwaite2010MagnetohydrodynamicMedium}. Though our findings are a step forward in constraining the thermal properties of X-ray cavities, further investigation will be required to fully characterize the material responsible for supporting them.

While CARMA has been decommissioned, new SZ instruments capable of observing the northern sky, such as NIKA2 and MUSTANG2, are well placed to make deep, spatially resolved measurements of the cavities in MS0735 at frequencies of 90-200 GHz. The vast majority of known cavity systems are however $\sim 2-10 ''$ in scale \citep{Hlavacek-Larrondo2015X-RAY1.2}, smaller than the $10'' - 20''$ resolution provided by these imaging cameras, though may be within the grasp of ALMA interferometric observations for some of these systems (for the smallest scale cavities, however, the lower brightness sensitivity at higher resolutions would make ALMA observations challenging). However, higher sensitivity alone may not be sufficient to distinguish between pressure support by diffuse, ultra-hot, thermal plasma and non-thermal particles. Multi-frequency observation, exploiting the spectral features of the SZ effect, may ultimately prove more valuable \citep{Colafrancesco2003TheGalaxies,Pfrommer2005UnveilingEffect,Colafrancesco2005TheCavities}, though current spectral studies of the kinematic SZ effect demonstrate the difficulty of high-sensitivity multi-frequency SZ observations (e.g., \citealt{Mroczkowski2012ABOLOCAM}). Both ultra-hot thermal plasma and relativistic electrons also produce hard X-ray signatures that may be accessible to the next generation of flagship X-ray telescopes. Until then, the SZ effect provides the most direct means for constraining the contents of X-ray cavities.

\acknowledgments{
We thank Adrian Vantyghem for supplying the latest \textit{Chandra} X-ray images and profiles of MS0735 presented in \cite{Vantyghem2014CyclingClusters}. We thank Laura Birzan and Brian McNamara for supplying the VLA observations of MS0735 presented in \cite{Birzan2008RadiativePower}. We thank Damiano Capriolli for his insightful comments. We thank the anonymous referee for their helpful comments.

Support for CARMA construction was derived from the Gordon and Betty Moore Foundation; the Kenneth T. and Eileen L. Norris Foundation; the James S. McDonnell Foundation; the Associates of the California Institute of Technology; the University of Chicago; the states of California, Illinois, and Maryland; and the National Science Foundation. CARMA development and operations were supported by NSF under a cooperative agreement and by the CARMA partner universities; the work at Chicago was supported by NSF grant AST-1140019. Additional support was provided by PHY-0114422.
}

%\begin{thebibliography}{}

\bibliographystyle{aasjournal.bst}
\bibliography{biblio.bib}

%\end{thebibliography}

\end{document}